\documentclass{article}

\usepackage{multirow,multicol,bbding,siunitx,colortbl}
\usepackage{amsthm}
\newtheorem{remark}{Remark}%

\usepackage[normalem]{ulem}

\usepackage{tikz,graphicx,threeparttable,amsmath,amssymb,float}
\graphicspath{{./}}
\usetikzlibrary{shapes.geometric, shapes.symbols, shapes.misc, positioning, arrows.meta, calc, decorations.pathreplacing, patterns,fit, backgrounds}
\usepackage{pgfplots}
\pgfplotsset{compat=1.18}
\usepackage{wasysym}
\newcommand{\fc}{\CIRCLE}      
\newcommand{\pc}{\LEFTcircle}  
\newcommand{\ec}{\Circle}  

\usepackage{tcolorbox}
\tcbuselibrary{skins}

\newtcolorbox{dashedblock}[1][]{
  enhanced,
  colback=white, colframe=white,
  borderline={0.8pt}{0pt}{black, dashed},
  sharp corners, boxrule=0pt,
  remember as=#1,
  left=8pt, right=8pt, top=6pt, bottom=6pt
}

\definecolor{cDS}{HTML}{5B4FCF}     
\definecolor{cGemF}{HTML}{0F9D78}   
\definecolor{cHope}{HTML}{E8703A}   
\definecolor{cClaude}{HTML}{B11226} 

\newcommand{\rZero}{0.6pt}    
\newcommand{\rClaudeS}{1.1pt} 
\newcommand{\rHopeA}{2.1pt}   
\newcommand{\rGemFA}{2.8pt}   
\newcommand{\rDS}{5pt}        
\definecolor{wrongred}{RGB}{200,30,30}
\definecolor{truthgreen}{RGB}{0,120,60}
\definecolor{missyellow}{RGB}{255,243,205}
 
\newcommand{\err}[2]{\textcolor{wrongred}{#1}\,(\textcolor{truthgreen}{#2})}

\usepackage{booktabs,array,xcolor}
\usepackage{tabularx}
\newcolumntype{L}{>{\raggedright\arraybackslash}X}
\definecolor{hdrblue}{HTML}{2C3E50}
\definecolor{highgreen}{HTML}{27AE60}
\definecolor{medamber}{HTML}{E67E22}
\definecolor{lowred}{HTML}{C0392B}
\newcommand{\conf}[1]{%
  \ifx#1H{\color{highgreen}\textbf{High}}%
  \else\ifx#1M{\color{medamber}\textbf{Med}}%
  \else{\color{lowred}\textbf{Low}}%
  \fi\fi}
  
\definecolor{inputblue}{HTML}{4A90D9}
\definecolor{inputbluebg}{HTML}{E8F0FA}
\definecolor{agentgreen}{HTML}{27AE60}
\definecolor{agentgreenbg}{HTML}{E8F8EF}
\definecolor{judgeamber}{HTML}{E67E22}
\definecolor{judgeamberbg}{HTML}{FDF2E5}
\definecolor{extractteal}{HTML}{16A085}
\definecolor{extracttealbg}{HTML}{E0F5F0}
\definecolor{kbpurple}{HTML}{8E44AD}
\definecolor{kbpurplebg}{HTML}{F0E6F6}
\definecolor{arrowgray}{HTML}{7F8C8D}
\definecolor{loopred}{HTML}{C0392B}
\definecolor{textdark}{HTML}{2C3E50}
\definecolor{quesblue}{HTML}{2980B9}
\definecolor{quesbluebg}{HTML}{E1EEF8}
\definecolor{resultred}{HTML}{C0392B}
\definecolor{resultredbg}{HTML}{FADBD8}
\definecolor{critgray}{HTML}{5D6D7E}
\definecolor{critgraybg}{HTML}{EAECEE}
\definecolor{inputblue}{HTML}{4A90D9}
\definecolor{inputbluebg}{HTML}{E8F0FA}
\definecolor{agentgreen}{HTML}{27AE60}
\definecolor{agentgreenbg}{HTML}{E8F8EF}
\definecolor{judgeamber}{HTML}{E67E22}
\definecolor{judgeamberbg}{HTML}{FDF2E5}
\definecolor{extractteal}{HTML}{16A085}
\definecolor{extracttealbg}{HTML}{E0F5F0}
\definecolor{kbpurple}{HTML}{8E44AD}
\definecolor{kbpurplebg}{HTML}{F0E6F6}
\definecolor{arrowgray}{HTML}{7F8C8D}
\definecolor{loopred}{HTML}{C0392B}
\definecolor{textdark}{HTML}{2C3E50}

\tikzset{
    pics/human/.style={code={
        \fill[inputblue!70] (0,0.22) circle (0.13);
        \fill[inputblue!70] (-0.18,-0.22) -- (-0.12,0.08) -- (0.12,0.08) -- (0.18,-0.22) -- cycle;
        \draw[inputblue!70, line width=0.7pt, line cap=round]
            (-0.22,-0.05) -- (-0.08,0.0) -- (0.08,0.0) -- (0.22,-0.05);
    }},
}

\usepackage[preprint]{staix_2026}
\makeatletter
\renewcommand{\@notice}{}
\makeatother

\usepackage[utf8]{inputenc}
\usepackage[T1]{fontenc}
\usepackage{hyperref}
\usepackage{url}
\usepackage{booktabs}
\usepackage{amsfonts}
\usepackage{nicefrac}
\usepackage{microtype}
\usepackage{xcolor}
\usepackage{graphicx}
\usepackage{amsmath,amssymb,amsthm}

\title{Systematic Literature Reviews With Two Multi-Agentic Systems And Human-In-The-Loop}

\author{%
  Zexin Ren \\
  George Washington University \\
  Department of Statistics \\
  \texttt{zxren10@gwu.edu}\\
  \And
  Zixuan Zhao \\
  George Washington University \\
  Department of Statistics \\
  \texttt{zixuan.zhao@gwmail.gwu.edu}
  \And
  Qiyun Li \\
  University of San Fransisco \\
  Department of Epidemiology and Biostatistics \\
  \texttt{qil016@ucsd.edu}\\\And
  Yawen Wu \\
  George Washington University \\
  Department of Statistics \\
  \texttt{yawen@gwu.edu}\\\And
  Lanjing Wang \\
  University of Washington \\
  Department of Biomedical Informatics and Medical Education  \\
  \texttt{lanjingw@uw.edu}\\\And
  Renjie Luo \\
  George Washington University \\
  Department of Statistics \\
  \texttt{rluo92@gwmail.gwu.edu}\\\And
  Yi Xu \\
  University of Melbourne \\
  School of Population and Global Health \\
  \texttt{yix11@student.unimelb.edu.au}\\\And
  Qing Guo \\
  HopeAI, Inc \\
  Statistical Innovation \\
  \texttt{Qing.Guo@hopeai.co}\\\And
  Jin Shi \\
  HopeAI, Inc \\
  Statistical Innovation \\
  \texttt{bubbles@hopeai.co}\\\And
  En Xie \\
  HopeAI,Inc \\
  Engineering Group \\
  \texttt{en@hopeai.co}\\\And
  Feifang Hu\footnote{Corresponding Author} \\
  George Washington University \\
  Department of Statistics \\
  \texttt{feifang@gwu.edu}\\\And
  Qian Shi \\
  Mayo Clinic \\
  Department of Quantitative Health Sciences \\
  \texttt{shi.qian2@mayo.edu}
}

\begin{document}

\maketitle

\begin{abstract}

Systematic literature review of clinical trials drives regulatory decision-making, but conventional screening and extraction are time-consuming, labor-intensive, and vulnerable to study selection bias. We propose two fit-to-purpose multi-agentic systems (MAS) for systematic literature review, with human-in-the-loop. The screening MAS uses multiple LLM agents with heterogeneous personas and multi-round cross-review, and uniformly improves accuracy over a single-LLM baseline. The extraction MAS combines standardization, an iterative correction loop, and retrieval-based context control to ensure accuracy and scalability. Both MAS are specifically designed to support Human-In-The-Loop which is essential for clinical decisions.  {The novelty of the proposed approach lies in the system architecture rather than in any single foundation tools: the system can naturally benefit from future improvements in the underlying tools, for instance, stronger LLM agents, retrieval engines, image recognition methods, etc.} As a real-world application, a published network meta-analysis is reproduced by the MAS. The result recovers all trials from the original study and identifies additional eligible trials missed by manual review, leading to updated clinical conclusions.

\end{abstract}





\section{Introduction}
Before initiating new randomized controlled trials (RCT), a comprehensive analysis of existing literature on similar trials, especially those involving the current standard of care, is crucial for sample size determination. The Food and Drug Administration (FDA) guideline \cite{guidelinefdameta} explicitly highlights the critical role of meta-analysis in drug safety, noting that \textit{``Regulatory decisions related to drug safety are generally taken after considering the totality of available evidence, which may include meta-analytic findings, as well as other factors...''} In practice, systematic literature reviews are time-consuming, labor intensive and may suffer from screening bias. In fact, it takes up to $12$ months on average per comprehensive systematic review \cite{borah2017analysis}. A conventional systematic review, or equivalently, meta-analysis, on RCTs consists of two major steps: a thorough screening of all relevant trials in the trial registry and a careful data extraction to retrieve relevant endpoint information. Manual screening, however, is prone to missed studies and inconsistent decisions, which contributes to study selection bias \citep{bannach2019machine,drucker2016research} and manual data extraction is subject to human error and inconsistency, leading to biased estimates \cite{jonnalagadda2015automating}.

Various tools have been
proposed for different purposes \citep{bannach2019machine,marshall2013tools,oami2025optimal} to speed up the process. One commonly discussed approach is the `Artificial Intelligence-assisted' approach, or more specifically, the utilization of Large Language Models (LLMs) (or equivalently `agents') in meta-analyses, see, e.g., \citep{de2023artificial, li2025enhancing, ge2024leveraging,li2024enhancing} and more. 

Among those lines of work, \cite{cao2025development, li2024enhancing} focus on prompt optimization, either through a well-designed prompt or a iterative prompt optimization method via self-reflecting on inefficient ones. However, the resulting quality depend on specific area of the research question and whether the prompt is tuned for this task, which may lack generalizability. Another line of work is on (Single) Agentic Systems (SAS), utilizing LLMs for each task. It has been discussed in both literature, e.g., \cite{li2025enhancing,chen2025large,cao2025automation,oami2025optimal}, and as industry products {\color{blue}Elicit.org, DistillerSR}. The conceptual workflow of SAS is presented in \eqref{eq:screening workflow baseline}:  
\begin{equation}\label{eq:screening workflow baseline}
    \text{User's Query}\mapsto\text{A screening LLM} \mapsto \text{An extraction LLM} \mapsto \text{Final data set}.
\end{equation}
One approach to use SAS is to feed full-text publications for each study into a long-context LLM, such as Claude Opus 4.7, and ask it to extract the required values. As shown in Table~\ref{tab:ipsos-oneshot-errors}, this strategy is difficult to scale for systematic reviews for three reasons: (1) token consumption grows rapidly with the number of publications and the complexity of extraction requirements; (2) incorrect extraction can lead to incorrect clinical decisions; and (3) LLM's stochasticity limits reproducibility, especially when no structured human-in-the-loop mechanism is available to audit uncertain cases. These limitations motivate our multi-agent design. For broader surveys of LLM-assisted SLR, see \citep{khalil2022tools,blaizot2022using,de2023artificial,ge2024leveraging,xu2025large,Li_Mathrani_Susnjak_2026}. A comparison of our method with some representative tools is summarized in Table~\ref{tab:tool-comparison}.

\begin{table*}[h]
\centering
\footnotesize
\caption{Capability comparison of the proposed system against some AI-assisted literature-review tools as of June 2026. \fc{}~=~fully supported; \pc{}~=~partial;
\ec{}~=~not supported or out of scope; \textbf{unc} ~=~ unclear. $\dag:$ Supports screening non-NCT registries, e.g., jRCT, ChiCTR, EudraCT, etc. $\S:$ Supports screening by major clinical conference, e.g., ASCO, AACR, ESMO, etc. }
\label{tab:tool-comparison}
\begin{tabular}{@{}p{2.8cm} *{5}{>{\centering\arraybackslash}p{1.6cm}}@{}}
\toprule
Capability & This work & Elicit & DistillerSR & Rayyan & ottoSR \\
\midrule
Screen by registry$^\dag$ & \fc & \ec & \ec & \ec & \ec \\
Screen by conference$^\S$ & \fc & \ec & \ec & \ec & \ec \\
Use of AI  & \fc & \fc & \fc & \fc & \fc \\
Multi-agentic & \fc & unc & unc & unc & unc \\
Data extraction & \fc & \fc & \fc & \fc  & \fc  \\
RoB assessment & \ec & \ec & \fc & \fc & \fc \\
Automation & \pc & \pc & \fc & \pc & \fc \\
Open-source & \pc & \pc & \ec & \ec & \ec \\
\bottomrule
\end{tabular}
\end{table*}

Despite the volume of research in this area, the practical adoption of LLMs for clinical decision-making in RCTs remains limited. In this paper, a large semi-automated, hierarchical Multi-Agent System (MAS), consisting of $2$ MAS with decentralized planning and decentralized execution
\cite{cheng2024exploring}, is proposed to screen for eligible studies and extract relevant endpoint information. 
Different from previous research that focuses on SAS, we propose to separate screening and extraction into two fit-to-purpose MAS and reserve spaces for \textbf{Human-In-The-Loop} for such tasks involving clinical decisions. {The novelty of our approach is \textbf{system-level}, thus any improvement in frontline LLM tools, e.g., the development of stronger foundation models, will result in the improvement of our system as well.}

Our approach offers several unique contributions. First, we decouple screening and data extraction into two fit-to-purpose MAS modules. For screening, the MAS uniformly improves accuracy and stability while pointing out ambiguous cases. For extraction, the MAS combines standardization, iterative correction, and retrieval-based context control to deliver credible extraction while keeping token usage under control. More importantly, it provides space for human audition. Together, these design choices support a reproducible, trustworthy workflow for systematic literature reviews of clinical trials. {Maybe the least significant advantage our workflow shares with other LLM-based screening methods is the substantial reduction in screening time \citep{borah2017analysis}. }

The rest of the paper is organized as follows. Section~\ref{sec:methods} introduces the proposed MAS. Section~\ref{sec:results} evaluates the method through screening and data-extraction experiments. Section~\ref{sec:adapting published systematic reviews} applies the system to a published network meta-analysis. Section~\ref{sec:conclusion} summarizes the paper. Prompts, supplementary evaluation results, and supporting network-meta-analysis materials are given in Appendices~\ref{app:screening-mas}--\ref{app:nma-replication}.

\section{Methods}\label{sec:methods}

\subsection{The Knowledge Base}\label{subsec:knowledge base}
The infrastructure of the workflow architecture is an establishment of clinical trial knowledge base. The knowledge base can be viewed as a local, dynamically updated repository that mirrors the relevant, up-to-date subset of major clinical trial registries. In this article, we use \textit{ClinicalTrials.gov} as the primary data source. The knowledge base stores structured trial metadata, including key fields such as “Study Overview,” “Intervention/Treatment,” “Eligibility Criteria,” “Outcome Measures,” and “Study Start Date.” It also stores online published articles for those trials.

\subsection{Multi-Agentic Screening with human in-the-loop}\label{subsec:screening}
In this subsection, we study an MAS for screening trials by NCT identifiers (NCTids). 

\begin{figure}[htbp]
    \centering
    \resizebox{\textwidth}{!}{
\begin{tikzpicture}[
    font=\sffamily\small,
    >=Stealth,
    base/.style={rounded corners=3pt, minimum height=8mm, align=center, text=textdark, font=\sffamily\footnotesize, inner sep=4pt, line width=0.6pt},
    input node/.style={base, fill=inputbluebg, draw=inputblue, minimum width=20mm, font=\sffamily\footnotesize\bfseries},
    kb node/.style={base, fill=kbpurplebg, draw=kbpurple, minimum width=22mm},
    agent a/.style={base, fill=agentgreenbg, draw=agentgreen, minimum width=15mm, minimum height=7.5mm},
    agent b/.style={base, fill=extracttealbg, draw=extractteal, minimum width=15mm, minimum height=7.5mm},
    judge node/.style={base, fill=judgeamberbg, draw=judgeamber, minimum width=16mm},
    crit node/.style={base, fill=critgraybg, draw=critgray, minimum width=16mm},
    ques node/.style={base, fill=quesbluebg, draw=quesblue, minimum width=15mm, minimum height=7.5mm},
    result node/.style={base, fill=resultredbg, draw=resultred, minimum width=24mm},
    flow/.style={-{Stealth[length=4.5pt, width=3.5pt]}, line width=0.6pt, color=arrowgray},
    feedback/.style={-{Stealth[length=4.5pt, width=3.5pt]}, line width=0.6pt, color=loopred!70, dashed},
    feedstub/.style={line width=0.6pt, color=loopred!70, dashed},
    grouplabel/.style={font=\sffamily\scriptsize\color{arrowgray}, align=center},
]
 
 
\node[input node] (query) at (0, 0) {User Query};
\node[input node] (proc)  at (0, -1.5) {Processing};
 
\node[kb node] (kb) at (2.8, -1.5) {Knowledge\\[-1pt]Base};
\begin{scope}[on background layer]
    \node[kb node, fill=kbpurplebg!60, draw=kbpurple!40, line width=0.4pt] at ($(kb)+(2pt,2pt)$) {};
    \node[kb node, fill=kbpurplebg!30, draw=kbpurple!25, line width=0.4pt] at ($(kb)+(4pt,4pt)$) {};
\end{scope}
 
\node[agent a] (a1) at (5.6, -0.6) {$A_1$};
\node[agent a] (a2) at (5.6, -1.5) {$A_2$};
\node[font=\normalsize, text=agentgreen!80] at (5.6, -2.2) (adots) {$\vdots$};
\node[agent a] (an) at (5.6, -2.8) {$A_N$};
\node[grouplabel, above=3pt of a1] {Screening\\[-1pt]Agents};
 
\node[judge node] (judge) at (8.0, -1.5) {$I$};
\node[grouplabel, above=3pt of judge] {Inspector};
\node[crit node] (crit)  at (8.0, -2.8) {Criteria};
 
\node[ques node] (q1) at (10.4, -0.6) {$Q_1$};
\node[ques node] (q2) at (10.4, -1.5) {$Q_2$};
\node[font=\normalsize, text=quesblue!80] at (10.4, -2.2) (qdots) {$\vdots$};
\node[ques node] (qn) at (10.4, -2.8) {$Q_N$};
\node[grouplabel, above=3pt of q1] {Follow-up\\[-1pt]Questions};
 
\node[agent b] (b1) at (2.8, -3.8) {$B_1$};
\node[agent b] (b2) at (2.8, -4.6) {$B_2$};
\node[font=\normalsize, text=extractteal!80] at (2.8, -5.2) (bdots) {$\vdots$};
\node[agent b] (bn) at (2.8, -5.8) {$B_{N^*}$};
\node[grouplabel, above=3pt of b1] {External\\[-1pt]Screening Agents};
 
\node[result node] (result) at (8.0, -4.2) {Result\\[-1pt]Agent};
 
\node (user) at (10.4, -4.2) {\tikz[scale=2.2]{\pic{human};}};
\node[font=\sffamily\scriptsize, text=textdark, below=2pt of user] {User};
 
\node[input node] (nctid) at (10.4, -5.8) {NCT ID\\[-1pt]List};
 
 
\draw[flow] (query) -- (proc);
 
\draw[flow] (proc) -- (kb);
 
\coordinate (kbout) at ($(kb.east)+(0.06,0)$);
\draw[flow] (kbout) -- ++(0.65,0) |- (a1.west);
\draw[flow] (kbout) -- ++(0.65,0) |- (a2.west);
\draw[flow] (kbout) -- ++(0.65,0) |- (an.west);
 
\draw[flow] (a1.east) -- ++(0.5,0) |- (judge.west);
\draw[flow] (a2.east) -- (judge.west);
\draw[flow] (an.east) -- ++(0.5,0) |- (judge.west);
 
\draw[flow] ($(judge.south)-(0.08,0)$) -- ($(crit.north)-(0.08,0)$);
\draw[flow] ($(crit.north)+(0.08,0)$) -- ($(judge.south)+(0.08,0)$);
 
\draw[flow] (crit.south) -- (result.north)
    node[midway, right=2pt, font=\sffamily\scriptsize] {Pass};
 
\draw[flow] (judge.east) -- ++(0.5,0) |- (q1.west);
\draw[flow] (judge.east) -- ++(0.5,0) |- (q2.west);
\draw[flow] (judge.east) -- ++(0.5,0) |- (qn.west);
 
\coordinate (trunkx) at ($(qn.east)+(0.5,0)$);
\draw[feedstub] (trunkx |- qn.east) -- (trunkx |- query);
\draw[feedstub] (q1.east) -- (trunkx |- q1.east);
\draw[feedstub] (q2.east) -- (trunkx |- q2.east);
\draw[feedstub] (qn.east) -- (trunkx |- qn.east);
\draw[feedback, rounded corners=4pt]
    (trunkx |- query) -- (a1.north |- query) -- (a1.north);
\node[font=\sffamily\scriptsize\itshape, text=loopred!70, above=2pt]
    at ($(trunkx |- query)!0.5!(a1.north |- query)$) {Feedback};
 
\coordinate (btrunk) at ($(proc.south)$);
\draw[line width=0.6pt, color=arrowgray, rounded corners=3pt] 
    (proc.south) -- (btrunk) -- (btrunk |- bn.west);
\draw[flow] (btrunk |- b1.west) -- (b1.west);
\draw[flow] (btrunk |- b2.west) -- (b2.west);
\draw[flow] (btrunk |- bn.west) -- (bn.west);
 
\draw[flow] (b1.east) -- ++(0.5,0) |- (result.west);
\draw[flow] (b2.east) -- ++(0.5,0) |- (result.west);
\draw[flow] (bn.east) -- ++(0.5,0) |- (result.west);
 
\draw[flow] (result.east) -- (user.west);
 
\draw[flow] (user.south) -- (nctid.north);
 
\end{tikzpicture}}
\caption{The proposed workflow for NCTid screening. The criteria block returns ``Pass'' if either the maximum turn, $T$, is reached or all agents return the same answer.}\label{fig:screening workflow W}
\end{figure}
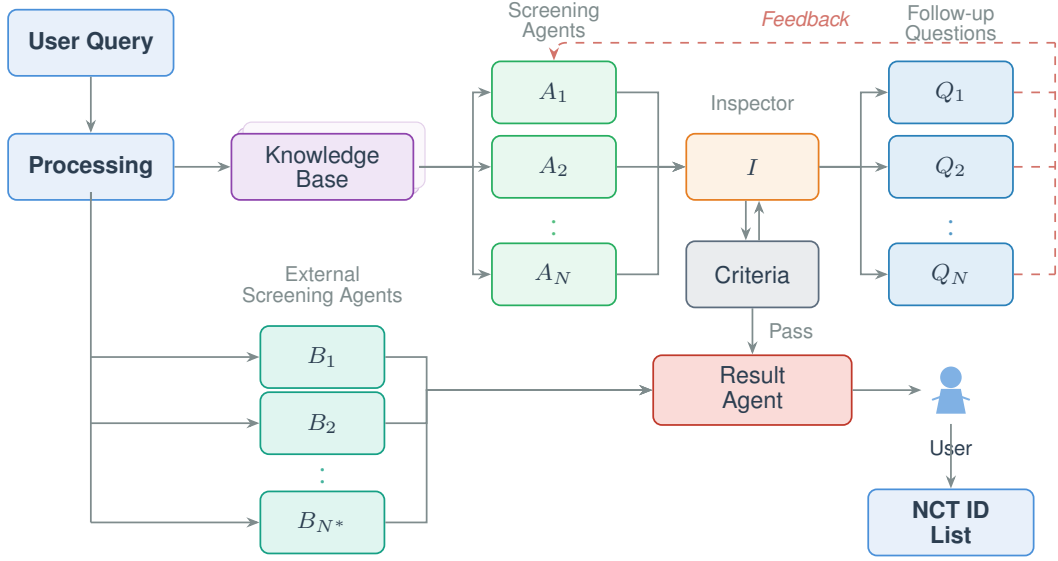

Figure~\ref{fig:screening workflow W} summarizes the screening MAS. A query-processing agent parses the user's eligibility criteria into structured Boolean inclusion/exclusion rules, which are broadcast to $N$ independent agents $A_1,\ldots, A_N$. Each agent labels every candidate trial as ``yes'', ``no'', or ``maybe'' with a brief rationale supporting its inclusion or exclusion decision.

An inspector agent $I$ checks agreement across the $N$ labels. If all agents agree, the decision is forwarded to the result agent. Otherwise, $I$ generates targeted follow-up questions $Q_1,\ldots,Q_N$ for each agent, initiating a new screening round; this self-revision repeats for at most $T$ rounds. In parallel, $N^*$ external deep-research agents $B_1,\ldots, B_{N^*}$ with web-search tools identify potentially eligible trials not registered with an NCTid (e.g., Japan/China/Korean clinical trial registries ChiCTR, UMIN, KCT, and so on).

\textbf{Human-in-the-loop.} Trials consistently with ``yes'' or ``no'' labels are finalized automatically. Trials that remain non-consensus after $T$ rounds, reach consensus on “maybe,” or are newly identified by external agents outside the original NCT-based candidate set are flagged for human review.

All agents are LLMs with task-specific prompts (see Appendix~\ref{app:screening-mas}). The MAS structure itself is independent of the LLM, so one can choose a variety of different LLMs as agent. To mitigate LLM mode collapse \cite{zhang2025verbalized}, the screening agents are equipped with heterogeneous personas; see Appendix~\ref{app:screening-prompts} for more discussions.

\subsection{Multi-Agentic Data Extraction with human-in-the-loop}\label{subsec:data extraction}

\textbf{Why a single LLM fails.} Table \ref{tab:ipsos-oneshot-errors} illustrates the failure modes of one-shot single-LLM extraction (Using Gemini 3.1 Pro to extract information from \cite{lee2023first}): only $55.6\%$ of OS-related cells are correct, with hallucinations, value substitutions, and missing rows. Stronger long-context models (e.g., Claude Opus 4.8) reduce errors but make token cost scale linearly with publication count, so brute-force long-context extraction does not scale to systematic reviews. The prompt that reproduces Table \ref{tab:ipsos-oneshot-errors} can be found in Appendix~\ref{app:sas-extraction-baseline}. In contrast, our extraction MAS with the same LLM model recovers all of Table \ref{tab:ipsos-oneshot-errors} correctly.

\begin{figure}[htbp]
    \centering
    \resizebox{\textwidth}{!}{
\begin{tikzpicture}[
    font=\sffamily\small,
    >=Stealth,
    base/.style={
        rounded corners=4pt,
        minimum height=10mm,
        align=center,
        text=textdark,
        font=\sffamily\footnotesize,
        inner sep=5pt,
    },
    input node/.style={
        base,
        fill=inputbluebg,
        draw=inputblue,
        line width=0.6pt,
        minimum width=26mm,
        font=\sffamily\footnotesize\bfseries,
    },
    agent node/.style={
        base,
        fill=agentgreenbg,
        draw=agentgreen,
        line width=0.6pt,
        minimum width=18mm,
    },
    judge node/.style={
        base,
        fill=judgeamberbg,
        draw=judgeamber,
        line width=0.6pt,
        minimum width=32mm,
    },
    extract node/.style={
        base,
        fill=extracttealbg,
        draw=extractteal,
        line width=0.6pt,
        minimum width=32mm,
    },
    kb node/.style={
        base,
        fill=kbpurplebg,
        draw=kbpurple,
        line width=0.6pt,
        minimum width=24mm,
    },
    flow/.style={
        -{Stealth[length=5pt, width=4pt]},
        line width=0.7pt,
        color=arrowgray,
    },
    loop/.style={
        -{Stealth[length=5pt, width=4pt]},
        line width=0.7pt,
        color=loopred!80,
        dashed,
    },
]
 
\node[agent node] (TA) {TA};
\node[agent node, below=0.7cm of TA] (SA) {SA};
\node[agent node, below=0.7cm of SA] (EA) {EA};
\node[font=\sffamily\scriptsize\color{arrowgray}, above=5pt of TA] {Specialized Agents};
 
\node[input node] at ($(TA.west)+(-2.5,0)$) (nct) {NCT IDs\\[-1pt]{\small\color{inputblue}(from screening)}};
\node[input node] at ($(EA.west)+(-2.5,0)$) (query) {User Query};
 
\node[kb node, above=1cm of TA] (kb) {Knowledge\\[-1pt]Base};
\begin{scope}[on background layer]
    \node[kb node, fill=kbpurplebg!60, draw=kbpurple!40,
          line width=0.4pt] at ($(kb)+(2pt,2pt)$) {};
    \node[kb node, fill=kbpurplebg!30, draw=kbpurple!25,
          line width=0.4pt] at ($(kb)+(4pt,4pt)$) {};
\end{scope}
 
\node[judge node, right=1.4cm of SA] (std) {Standardize\\[-1pt]Agent};
 
\node[extract node, right=1.4cm of std] (ext) {Extraction\\[-1pt]Agent};

\node[extract node, below=0.5cm of ext](retrieve) {Retrieval\\[-1pt]Context Control};
 
\node[right=0.5cm of ext, inner sep=0pt] (table) {%
    \footnotesize\sffamily
\setlength{\tabcolsep}{3pt}
\begin{tabular}{lllcccc}
\toprule
\textcolor{hdrblue}{\textbf{NCT ID}} & \textcolor{hdrblue}{\textbf{Arm}} & \textcolor{hdrblue}{\textbf{Subgroup}} & \textcolor{hdrblue}{\textbf{EP}} & \textcolor{hdrblue}{\textbf{HR}} & \textcolor{hdrblue}{\textbf{95\% CI}} & \textcolor{hdrblue}{\textbf{Conf.}} \\
\midrule
NCT001 & Exp.  & PD-L1$\geq$1\% & PFS & 0.72 & (0.58, 0.89) & \conf{H} \\
NCT001 & Ctrl. & PD-L1$\geq$1\% & PFS & ---  & ---           & \conf{H} \\
NCT002 & Exp.  & PD-L1$<$1\%    & OS  & 0.81 & (0.66, 0.99) & \conf{M} \\
NCT003 & Exp.  & PD-L1$\geq$1\% & PFS & 0.65 & (0.50, 0.84) & \conf{L} \\
\bottomrule
\end{tabular}
};
 
\draw[flow] (nct.east) -- ++(0.55,0) |- (TA.west);
\draw[flow] (nct.east) -- ++(0.55,0) |- (SA.west);
\draw[flow] (nct.east) -- ++(0.55,0) |- (EA.west);
\draw[flow] (query.east) -- ++(0.55,0) |- (TA.west);
\draw[flow] (query.east) -- ++(0.55,0) |- (SA.west);
\draw[flow] (query.east) -- ++(0.55,0) |- (EA.west);
 
\draw[flow] (TA.east) -- ++(0.45,0) |- (std.west);
\draw[flow] (SA.east) -- (std.west);
\draw[flow] (EA.east) -- ++(0.45,0) |- (std.west);
 
\draw[flow] (std.east) -- (ext.west);
\draw[flow] (ext.east) -- (table.west);
\draw[flow] (retrieve.north) -- (ext.south);
\draw[flow, color=kbpurple!60] (kb.south) -- (TA.north);
\draw[flow, color=kbpurple!60] (kb.east) -| (std.north);

\draw[loop] (ext) to[out=50, in=130, looseness=3.5, min distance=15mm] (ext);
\node[font=\sffamily\scriptsize\itshape, text=loopred!80, 
      above=26pt of ext] {\textbf{Repeat $T$ rounds}};
 
\end{tikzpicture}}
\caption{The proposed workflow for data extraction. TA, SA, EA represent treatment arm agent, subgroup agent, endpoints agent, respectively.}\label{fig:extraction workflow}
\end{figure}
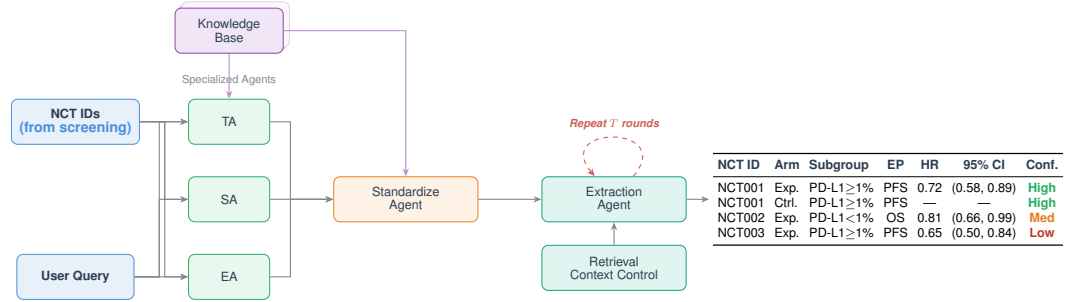

\begin{table}[htbp]
\centering
\caption{One-shot LLM extraction of IPSOS trial efficacy outcomes by PD-L1 status. Of $12$ reportable subgroup rows, $8$ contained at least one extraction error and $2$ rows are omitted, illustrating the instability of one-shot extraction for SLR data capture.}
\label{tab:ipsos-oneshot-errors}
\resizebox{\textwidth}{!}{%
\begin{tabular}{llllrrrrrrrrr}
\toprule
\multirow{2}{*}{Study ID} & \multirow{2}{*}{Treatment} & \multirow{2}{*}{Drugs} & \multirow{2}{*}{PD-L1 Status} &
\multicolumn{3}{c}{Median PFS (Months)} &
\multicolumn{3}{c}{Median OS (Months)} &
\multicolumn{3}{c}{ORR (\%)} \\
\cmidrule(lr){5-7} \cmidrule(lr){8-10} \cmidrule(lr){11-13}
 & & & & Median & Lower 95\% CI & Upper 95\% CI & Median & Lower 95\% CI & Upper 95\% CI & ORR & Lower 95\% CI & Upper 95\% CI \\
\midrule
IPSOS & Experimental & Atezolizumab               & -- (ITT)        & 4.2 & 3.7 & 5.5 & 10.3            & 9.4              & 11.9             & 17\%  & 12.8\% & 21.6\% \\
IPSOS & Control      & Vinorelbine or Gemcitabine & -- (ITT)        & 4.0 & 2.9 & 5.4 & 9.2             & 5.9              & 11.2             & 8\%   & 4.2\%  & 13.5\% \\
IPSOS & Experimental & Atezolizumab               & TC $<$1\%       & NR  & NR  & NR  & 10.3            & \err{7.1}{NR}    & \err{14.3}{NR}   & NR    & NR     & NR     \\
IPSOS & Control      & Vinorelbine or Gemcitabine & TC $<$1\%       & NR  & NR  & NR  & \err{10.2}{7.1} & \err{5.4}{NR}    & \err{12.1}{NR}   & NR    & NR     & NR     \\
IPSOS & Experimental & Atezolizumab               & TC $\geq$1\%    & NR  & NR  & NR  & \err{10.3}{9.4} & \err{8.4}{NR}    & \err{13.9}{NR}   & NR    & NR     & NR     \\
IPSOS & Control      & Vinorelbine or Gemcitabine & TC $\geq$1\%    & NR  & NR  & NR  & \err{9.4}{10.3} & \err{5.5}{NR}    & \err{10.3}{NR}   & NR    & NR     & NR     \\
IPSOS & Experimental & Atezolizumab               & TC 1--49\%      & NR  & NR  & NR  & \err{10.2}{8.4} & NR               & NR               & NR    & NR     & NR     \\
IPSOS & Control      & Vinorelbine or Gemcitabine & TC 1--49\%      & NR  & NR  & NR  & \err{8.4}{10.2} & NR               & NR               & NR    & NR     & NR     \\
IPSOS & Experimental & Atezolizumab               & TC $\geq$50\%   & NR  & NR  & NR  & 11.0            & NR               & NR               & NR    & NR     & NR     \\
IPSOS & Control      & Vinorelbine or Gemcitabine & TC $\geq$50\%   & NR  & NR  & NR  & \err{7.7}{13.9} & NR               & NR               & NR    & NR     & NR     \\
\rowcolor{missyellow} IPSOS & Experimental & Atezolizumab               & Unknown & NR & NR & NR & 16.9 & NR & NR & NR & NR & NR \\
\rowcolor{missyellow} IPSOS & Control      & Vinorelbine or Gemcitabine & Unknown & NR & NR & NR & 9.4  & NR & NR & NR & NR & NR \\
\bottomrule
\end{tabular}%
}
\par\vspace{0.4em}
\footnotesize
\textcolor{wrongred}{Red} = value returned by the one-shot LLM extraction;
\textcolor{truthgreen}{green} = correct value from the primary publication;
\colorbox{missyellow}{yellow row} = subgroup the LLM misses to extract. CI, confidence interval; ITT, intent-to-treat; NR, not reported; ORR, objective response rate; OS, overall survival; PFS, progression-free survival; TC, tumor cell.
\end{table}
Given the screened NCTid list in Subsection \ref{subsec:screening} and a new user query specifying extraction details, the extraction MAS (Figure \ref{fig:extraction workflow}) proceeds in three steps: standardization, iterative extraction, and retrieval-based context control.

\begin{enumerate}
    \item \textbf{Standardization.} The first challenge is that the target extraction table is often undetermined before reviewing the publications. This is almost always the case because subgroup definitions may vary across studies or even across multiple papers from the same study. Three specialized agents—a treatment agent (TA), subgroup agent (SA), and endpoint agent (EA)—search the associated publications for all mentions of the query inputs. A standardization agent then unifies the heterogeneous names into a mock table with pre-specified columns. We provide one mock table column names as an example here: ``\textit{NCTid; study popular name; treatment name; arm type (experimental vs control); Cohort; Analysis Population (ITT vs per-Protocol vs Subgroup); Subgroup names; Sample size; Efficacy value; lower $95\%$ confidence interval; Upper $95\%$ confidence interval.}''
    \item \textbf{Iterative extraction.} An extraction agent fills the table by first proposing initial values, then re-reading the source PDF to correct them. This correction loop repeats for $T$ rounds, after which each cell receives a confidence label (High, Medium, or Low, see Appendix~\ref{app:extraction-confidence} for definitions) based on cross-round consistency. 
    \item \textbf{Retrieval-based context control.} A na\"ive implementation of extraction would pass all publications to the extraction agent in every round and for every table cell, which scales poorly in token cost. In our approach, source documents are automatically partitioned into evidence clips, such as texts, tables and figures. Each clip is indexed by structured attributes. When the extraction agent fills a specific cell, only clips matching the corresponding attributes are retrieved as input. The retrieval layer is implemented using OpenViking~\cite{volcengine2025openviking}. This design reduces irrelevant context.
\end{enumerate}

\textbf{Human-in-the-loop and source-risk errors.} In the MAS, Low- and Medium-confidence results are flagged directly, typically representing no more than 10\% of all extracted data points. High-confidence results are additionally flagged under two predefined source-risk conditions: (1) low-quality source material, including low-resolution figures (e.g., Kaplan-Meier plots or forest plots) or compressed supplementary files; and (2) publication-version ambiguity, including unclear data cut-off times or inconsistent values between primary and long-term follow-up reports. In practice, most inaccurate extractions were captured by these review flags, suggesting that the workflow substantially narrows the scope of manual correction while retaining human oversight.

Together, these three components address the principal SAS failure modes: missed information (resolved by iterative extraction with confidence labeling), hallucinated values (resolved by correction of pseudo-answers rather than regeneration \citep{zhang2025recursivelanguagemodels} and iterative extraction), and non-standardized output (resolved by the standardization step), while retrieval-based control keeps token cost tractable.

\subsection{Parallel screening}
The huge time cost of producing reliable meta-analyses is well-documented in the literature \cite{allen1999estimating,borah2017analysis}. The primary difficulty lies in the manual screening of large trial registries, where researchers must balance speed with the risk of exclusion errors. However, our workflow alters this constraint through parallelization. While our notation $A_1,\ldots,A_N$ implies individual agents, the architecture operationally functions as a distributed array of agentic reviewers, capable of processing the entire Knowledge Base concurrently. Similarly for data extraction. This parallel construction decouples the total processing time from the number of studies; a task that would require $5000$ minutes linearly (e.g., $5$ minutes per paper for $1000$ papers) is theoretically compressed to the latency of a single inference cycle, i.e., in total $5$ minutes. Given that this workflow trivializes the temporal burden, the subsequent sections focuses exclusively on the critical remaining difficulty: the accuracy and stability of the screening decisions.

\section{Results}\label{sec:results}

In this section, we evaluate the proposed workflow through large-scale screening and data-extraction experiments. Throughout the paper, we consider a wide range of models including Gemini 3.1 Pro Preview, ChatGPT 5.1, DeepSeek Reasoning, Claude Opus 4.8 Reasoning, Gemini 3 Flash Preview, and HopeAI Gemma4,\footnote{HopeAI Gemma4 is a customized LLM based on Gemma 4.} denoted as Gem, GPT, DS, Claude, GemF, and Hope, respectively. To control token consumption, the two large-scale evaluations used DS, GemF, and Hope as primary agent models, whereas the real-world application in Section~\ref{sec:adapting published systematic reviews} uses Gem, GPT, and Claude.

\subsection{Evaluation of the multi-agentic screening: accuracy and stability}\label{subsec:workflow evaluation}

In order to validate accuracy of the screening MAS, we conduct a large scale screening of first line non-small cell lung cancer (NSCLC)\footnote{The method is not NSCLC-specific and can be extended to other disease indications.} trials. In all, $2245$ trials \footnote{Selected by filter ``Phase2/Phase3+Interventional+Industry Sponsored'' on \textit{Clinicaltrials.gov}} from \textit{Clinicaltrials.gov} are included for screening. Furthermore, the trials are screened by the inclusion/exclusion criteria provided in Appendix~\ref{app:screening-nsclc}.Among all trials, $213$ are marked to satisfy the eligibility conditions by three independent reviewers blinded to the model outputs (Fleiss’ Kappa$=0.70$, $2$ sided $p$-value $0.00$). The reviewers discuss the results and reach consensus across all trials. Then, this final result serves as the \textbf{correct benchmark} to be compared with.

To control token consumption in this large-scale experiment, we used three relatively cost-efficient base models: DS, GemF, and Hope. For each base model, we compared the proposed MAS with its corresponding SAS baseline ($N=T=1$). The MAS used $N=5$ agents based on the same underlying model, with distinct personas introduced through prompting, and allowed up to $T=3$ revision rounds; see Appendix~\ref{app:screening-prompts}. For example, the DS-based MAS was compared with a single DS agent, so that performance differences could be attributed to the MAS workflow rather than to the models themselves. Inspector and summary agents were implemented using Claude. All models were deployed with $\texttt{temp}=0.7$ and $\texttt{top\_p}=0.9$. Because the workflow is semi-automated, all flagged cases were adjudicated by an independent reviewer, who recorded the final decision and additional review time.\footnote{External agents were not used in validating the screening MAS.}

Table~\ref{tab:ensemble_metrics} summarizes screening performance across 30 independent Monte Carlo replications on the NSCLC benchmark set, including accuracy, sensitivity, specificity, positive predictive value (PPV), F1 score, Matthews correlation coefficient (MCC), work saved over sampling (WSS), and the number of trials requiring human review. Sensitivity and PPV are particularly important in systematic literature reviews because missed eligible and/or included ineligible trials can distort the decisions. The single-agent models show distinct error profiles. In particular, the single DS agent is overly conservative, achieving near-perfect specificity and PPV but low sensitivity. The MAS corrects this failure mode: DS sensitivity increases from $0.583$ to $0.926$ while PPV remained high at $0.987$. GemF and Hope had stronger single-agent sensitivity, but MAS still improved their F1 scores. Overall, the results indicate that the MAS can reduce model-specific screening errors while maintaining high precision. The agent-disagreement mechanism also provides a principled way for human-in-the-loop, with the average number of cases requiring review depending on the base model. The cost-effectiveness plot is presented in Figure~\ref{fig:cost-effective}, using F1 score as the vertical axis.

Further in Figure~\ref{fig:pairwise_kappa} in Appendix~\ref{app:screening-nsclc}, the pairwise Cohen's Kappa are reported for the first $10$ runs. The average Cohen's Kappa is above $0.95$, indicating almost perfect agreement across runs. This reflects the stability of our screening MAS. In addition, we also conduct a smaller scale simulation based approach similar to \cite{li2024enhancing} on colorectal cancer (CRC) with more models. See Appendix~\ref{app:crc-screening-evaluation} for details.

\begin{table}[t!]
\centering
\caption{Performance of 30-run Monte Carlo screening systems on the
NSCLC benchmark ($n = 2236$), grouped by base model. Values are
mean (SD) across the 30 runs. \textbf{Bold} indicates that the
agentic approach strictly outperforms its own single-model baseline
on that metric. Rule~A: non-consensus resolved by majority votes from agents,
``Maybe'' resolved by human-in-the-loop. Rule~B: ``Maybe'' and
non-consensus resolved by human-in-the-loop. Token: total token consumption, input + output in millions, by the models.}
\label{tab:ensemble_metrics}
\small
\setlength{\tabcolsep}{4pt}
\resizebox{\textwidth}{!}{\begin{tabular}{llccccccccc}
\toprule
Model & Variant & Acc & Sens & Spec & PPV & F1 & MCC & WSS & \# Human Review & Token \\
\midrule
\multirow{3}{*}{DS}
  & Single    & 0.960 (0.002) & 0.583 (0.015) & 1.000 (0.000) & 0.997 (0.005) & 0.736 (0.013) & 0.746 (0.012) & 0.528 (0.014) & 150.3 (3.512) & 4.1m\\
  & Agentic-A & \textbf{0.975} (0.002) & 0.790 (0.016) & 0.995 (0.001) & 0.941 (0.008) & \textbf{0.859} (0.012) & \textbf{0.849} (0.012) & \textbf{0.710} (0.015) & \textbf{12.67} (3.606) & 17.7m\\
  & Agentic-B & \textbf{0.992} (0.001) & \textbf{0.926} (0.015) & 0.999 (0.001) & 0.987 (0.005) & \textbf{0.955} (0.007) & \textbf{0.951} (0.008) & \textbf{0.837} (0.013) & 150 (5.6) &17.7m\\
\midrule
\multirow{3}{*}{GemF}
  & Single    & 0.969 (0.000) & 0.991 (0.000) & 0.967 (0.000) & 0.755 (0.003) & 0.857 (0.002) & 0.850 (0.002) & 0.866 (0.000) & 0 (0) & 4.3m \\
  & Agentic-A & \textbf{0.970} (0.001) & 0.988 (0.003) & \textbf{0.968} (0.001) & \textbf{0.762} (0.006) & \textbf{0.861} (0.003) & \textbf{0.863} (0.003) & 0.865 (0.003) & 0 (0.5) &22.8m\\
  & Agentic-B & \textbf{0.973} (0.001) & \textbf{0.995} (0.000) & \textbf{0.971} (0.000) & \textbf{0.780} (0.007) & \textbf{0.874} (0.004) & \textbf{0.867} (0.004) & \textbf{0.874} (0.001) & 44 (4.7)& 22.8m\\
\midrule
\multirow{3}{*}{Hope}
  & Single    & 0.966 (0.000) & 0.941 (0.000) & 0.968 (0.000) & 0.755 (0.003) & 0.839 (0.003) & 0.826 (0.003) & 0.825 (0.000) & 0 (0) & 3.8m\\
  & Agentic-A & \textbf{0.967} (0.000) & 0.941 (0.001) & \textbf{0.974} (0.001) & \textbf{0.765} (0.003) & \textbf{0.844} (0.003) & \textbf{0.831} (0.003) & 0.824 (0.004) & 0 (0) & 20.3m \\
  & Agentic-B & \textbf{0.973} (0.001) & \textbf{0.974} (0.001) & \textbf{0.971} (0.000) & \textbf{0.796} (0.006) & \textbf{0.869} (0.004) & \textbf{0.859} (0.005) & \textbf{0.844} (0.005) & 25 (2.3)& 20.3m \\
  \midrule
  Claude & Single & 0.965 (0.000) & 0.981 (0.001)& 0.963 (0.003) & 0.736 (0.005) & 0.841 (0.010) & 0.832 (0.002) & 0.855 (0.005) & 7 (1.8)& 3.1m\\
\bottomrule
\end{tabular}}
\par\smallskip
\footnotesize Acc: accuracy; Sens: sensitivity; Spec: specificity;
PPV: positive predictive value; F1: F1 score; MCC: Matthews
correlation coefficient; WSS: work saved over sampling. For
\# Human Review, lower is better.
\end{table}

\begin{figure}[t!]
    \centering
    \begin{tikzpicture}
\begin{axis}[
    width=8.6cm, height=6.8cm,
    xmode=log, log basis x=10,
    xlabel={Total inference cost (USD, log scale)},
    ylabel={F1 score},
    xmin=0.3, xmax=80,
    ymin=0.70, ymax=1.00,
    xtick={0.5,1,2,5,10,20,50},
    xticklabels={\$0.5,\$1,\$2,\$5,\$10,\$20,\$50},
    ytick={0.70,0.75,0.80,0.85,0.90,0.95,1.00},
    grid=major,
    major grid style={line width=.2pt, draw=gray!22},
    xminorticks=false,
    tick align=outside,
    axis line style={gray!55},
    every axis label/.append style={font=\small},
    tick label style={font=\footnotesize},
    legend cell align=left,
    legend columns=3,
    legend style={
        at={(0.5,-0.26)}, anchor=north,
        draw=gray!45, fill=white, fill opacity=0.92, text opacity=1,
        font=\scriptsize, row sep=1pt, column sep=8pt,
        inner xsep=2pt, inner ysep=2pt,
        /tikz/every even column/.append style={column sep=8pt},
    },
]

\draw[-{Stealth[length=2.6mm,width=2mm]}, cHope, line width=1pt,
      dash pattern=on 4pt off 2.5pt, shorten <=2.5pt, shorten >=5pt]
      (axis cs:0.38,0.839) -- (axis cs:2.03,0.869);
\addplot[only marks, mark=o, mark size=\rZero, line width=1pt, color=cHope, forget plot]
      coordinates {(0.38,0.839)};
\addplot[only marks, mark=diamond*, mark size=\rHopeA, color=cHope, forget plot]
      coordinates {(2.03,0.869)};

\draw[-{Stealth[length=2.6mm,width=2mm]}, cDS, line width=1pt,
      dash pattern=on 4pt off 2.5pt, shorten <=7pt, shorten >=8pt]
      (axis cs:1.92,0.736) -- (axis cs:7.83,0.955);
\addplot[only marks, mark=o, mark size=\rDS, line width=1pt, color=cDS, forget plot]
      coordinates {(1.92,0.736)};
\addplot[only marks, mark=diamond*, mark size=\rDS, color=cDS, forget plot]
      coordinates {(7.83,0.955)};

\draw[-{Stealth[length=2.6mm,width=2mm]}, cGemF, line width=1pt,
      dash pattern=on 4pt off 2.5pt, shorten <=2.5pt, shorten >=6pt]
      (axis cs:11.28,0.857) -- (axis cs:59.6,0.874);
\addplot[only marks, mark=o, mark size=\rZero, line width=1pt, color=cGemF, forget plot]
      coordinates {(11.28,0.857)};
\addplot[only marks, mark=diamond*, mark size=\rGemFA, color=cGemF, forget plot]
      coordinates {(59.6,0.874)};

\addplot[only marks, mark=o, mark size=\rClaudeS, line width=1pt, color=cClaude, forget plot]
      coordinates {(18.2,0.841)};

\addlegendimage{cDS, line width=2pt}\addlegendentry{DS}
\addlegendimage{cGemF, line width=2pt}\addlegendentry{GemF}
\addlegendimage{cHope, line width=2pt}\addlegendentry{Hope}
\addlegendimage{cClaude, line width=2pt}\addlegendentry{Claude}
\addlegendimage{only marks, mark=o, mark size=2pt, line width=0.9pt, black}\addlegendentry{Single LLM}
\addlegendimage{only marks, mark=diamond*, mark size=2pt, black}\addlegendentry{Agent\,+\,Human}
\end{axis}
\end{tikzpicture}
    \caption{Cost-effectiveness of single-LLM versus agentic (agent\,+\,human-in-the-loop) screening on the NSCLC benchmark. The horizontal axis is the \emph{total inference cost} in USD on a log scale, obtained from the input/output token consumption in Table~\ref{tab:ensemble_metrics} priced at each model's published API rate. Arrows trace the single$\,\rightarrow\,$agentic trajectory for each model. Marker area is proportional to the number of records left for human review. Upper left is better.}
    \label{fig:cost-effective}
\end{figure}
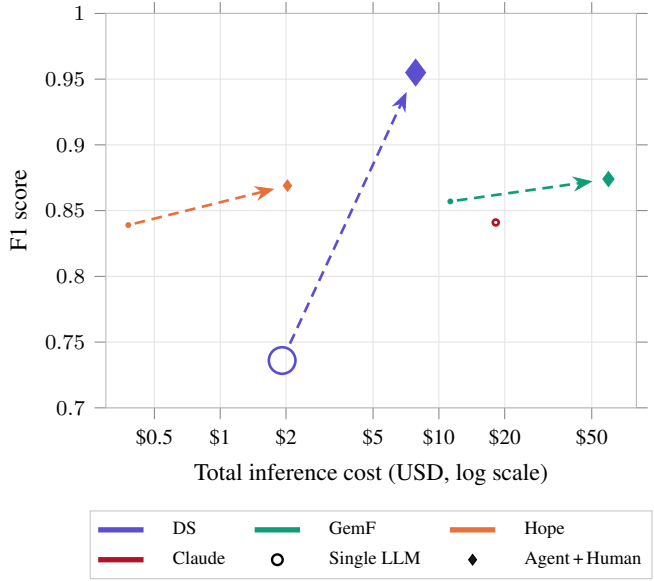

\subsection{Evaluation of multi-agentic data extraction}

We evaluated data-extraction accuracy across three disease indications: triple-negative breast cancer (TNBC), non-small-cell lung cancer (NSCLC), and gastric cancer (GC). For each indication, 10 phase 2 and/or phase 3 trials were randomly selected from ClinicalTrials.gov. For each trial, the workflow extracted median PFS, median OS, and objective response rate (ORR), together with their 95\% confidence intervals, for both the intention-to-treat population and PD-L1-defined subgroups. An extraction is scored as correct only if the reported point estimate, lower 95\% confidence limit, and upper 95\% confidence limit exactly matches the source publication. Unlike the controlled comparison in Section~\ref{subsec:workflow evaluation}, this evaluation uses three different foundation models—GemF, Hope, and DS—to extract together, with $3$ rounds of correction. The objective was to assess the feasibility of large-scale structured extraction.

Table~\ref{tab:extraction_accuracy} shows that the confidence labels provide useful risk stratification. Of 804 extracted entries, $741$ ($92.2\%$) were labeled as high confidence, of which $670$ ($90.4\%$) exactly matched the source publication. Incorrect high-confidence entries were predominantly attributable to the two source-risk patterns in Section~\ref{subsec:data extraction}. With human review of source-risk-flagged high-confidence cells, the extraction accuracy for high-confidence cells rises up to $97\%$. Further, if all low/medium-confidence cells ($63$ entries, $7.8\%$) are also revised the overall extraction accuracy rises to $97\%$, with $783$ out of $804$ correct.

\begin{table}[!htbp]
\centering
\caption{Extraction accuracy of the data-extraction MAS across three disease indications, three endpoints, and two populations (ITT: intention-to-treat; mPFS/mOS: median progression-free/overall survival; ORR: overall response rate). An extraction is scored correct if and only if all components of the reported quantity (point estimate, lower 95\% CI bound, upper 95\% CI bound) exactly match the source publication. Confidence labels follow the rule in Appendix~\ref{app:extraction-confidence}.}
\label{tab:extraction_accuracy}

\begin{threeparttable}[t!]
\footnotesize
\setlength{\tabcolsep}{3pt}
\renewcommand{\arraystretch}{1.0}
\begin{tabular}{@{}l l l c c c c c c@{}}
\toprule
\multirow{2}{*}{\textbf{Indication}} &
\multirow{2}{*}{\textbf{Endpoint}} &
\multirow{2}{*}{\textbf{Population}} &
\multicolumn{3}{c}{\textbf{Accuracy by confidence level\tnote{a}}} &
\multirow{2}{*}{\textbf{Overall}} & \textbf{{Source-risk/LLM error\tnote{c}}} &\textbf{{Token cost}} \\
\cmidrule(lr){4-6}
 & & & \textbf{High} & \textbf{Medium} & \textbf{Low} & & {\textbf{(Acc. after correction)\tnote{c}}}& {(\textbf{m: millions})}\\
\midrule
\multirow{7}{*}{TNBC}
 & \multirow{2}{*}{mPFS}
   & ITT                & 16/16 (100.0\%) & --/0  & 0/1  & 16/17 & 0/0 (100.0\%) &  \multirow{6}{*}{TA/SA/EA:7.7m}\\
 & & Subgroup\tnote{b}  & 33/41 (80.5\%)  & --/0  & 0/3  & 33/44 & 6/2 (95.1\%)& \\
 & \multirow{2}{*}{mOS}
   & ITT                & 15/15 (100.0\%) & 1/1   & 0/1  & 16/17 & 0/0 (100.0\%)& \\
 & & Subgroup           & 38/41 (92.7\%)  & 1/2   & 1/1  & 40/44 & 2/1 (97.6\%)&\multirow{4}{*}{Extraction:18.5m}\\
 & \multirow{2}{*}{ORR}
   & ITT                & 15/16 (93.8\%)  & --/0  & 0/1  & 15/17 &0/1 (93.8\%)& \\
 & & Subgroup           & 41/42 (97.6\%)  & 0/1   & 0/1  & 41/44 &1/0 (100\%)& \\
\cmidrule(l){2-8}
 & \multicolumn{2}{l}{\textit{TNBC subtotal}} & \textit{158/171 (92.4\%)} & \textit{2/4} & \textit{1/8} & \textit{161/183} & \textit{9/4 (\textbf{96.5\%})}  & \\
\midrule
\multirow{7}{*}{NSCLC}
 & \multirow{2}{*}{mPFS}
   & ITT                & 17/18 (94.4\%)  & 1/2   & 0/1  & 18/21 & 0/1 (94.4\%)& \multirow{6}{*}{TA/SA/EA:11.8m}\\
 & & Subgroup           & 71/76 (93.4\%)  & --/0  & 0/3  & 71/79  & 5/0 (100\%) & \\
 & \multirow{2}{*}{mOS}
   & ITT                & 17/19 (89.5\%)  & 0/1   & 0/1  & 17/21 & 2/0 (100\%)& \\
 & & Subgroup           & 66/75 (88.0\%)  & 0/3   & 0/1  & 66/79 & 6/3 (96.0\%)& \multirow{4}{*}{Extraction: 30.3m}\\
 & \multirow{2}{*}{ORR}
   & ITT                & 18/21 (85.7\%)  & --/0  & --/0 & 18/21 &2/1 (95.2\%)& \\
 & & Subgroup           & 70/77 (90.9\%)  & 0/1   & 0/1  & 70/79 &4/3 (96.1\%)&  \\
\cmidrule(l){2-8}
 & \multicolumn{2}{l}{\textit{NSCLC subtotal}} & \textit{259/286 (90.6\%)} & \textit{1/7} & \textit{0/7} & \textit{260/300}  & \textit{19/8 (\textbf{97.0\%})}& \\
\midrule
\multirow{7}{*}{GC}
 & \multirow{2}{*}{mPFS}
   & ITT                & 15/16 (93.8\%)  & 1/2   & 0/2  & 16/20 & 1/0 (100\%)& \multirow{6}{*}{TA/SA/EA:12.2m}\\
 & & Subgroup           & 67/76 (88.2\%)  & 3/5   & 3/6  & 73/87 &8/1 (98.6\%)& \\
 & \multirow{2}{*}{mOS}
   & ITT                & 15/15 (100.0\%) & 0/2   & 1/3  & 16/20 &0/0 (100\%)& \\
 & & Subgroup           & 66/80 (82.5\%)  & 1/3   & 2/4  & 69/87 & 8/6 (92.5\%)& \multirow{4}{*}{Extraction:29.9m}\\
 & \multirow{2}{*}{ORR}
   & ITT                & 14/15 (93.3\%)  & 3/4   & 0/1  & 17/20 &1/0 (100\%)& \\
 & & Subgroup           & 76/82 (92.7\%)  & 1/2   & 1/3  & 78/87 &4/2 (97.6\%) & \\
\cmidrule(l){2-8}
 & \multicolumn{2}{l}{\textit{GC subtotal}} & \textit{253/284 (89.1\%)} & \textit{9/18} & \textit{7/19} & \textit{269/321} &\textit{22/9 (\textbf{96.8\%})}& \\
\midrule
\multicolumn{2}{l}{Pooled} & ITT      & 142/151 (94.0\%)         & 6/12            & 1/11          & 149/174 &6/3 (98.0\%)& \\
\multicolumn{2}{l}{Pooled} & Subgroup & 528/590 (89.5\%)         & 6/17            & 7/23          & 541/630 & 44/18 (96.9\%) & \\
\multicolumn{2}{l}{Pooled} & All      & \textbf{670/741 (90.4\%)} & \textbf{12/29} & \textbf{8/34} & \textbf{690/804} &50/21 (\textbf{97.2\%})& \\
\bottomrule
\end{tabular}

\begin{tablenotes}[flushleft]\scriptsize
\item[a] Entries are (correct extractions)/(all extractions), counting efficacy values and 95\% CIs.
\item[b] Any PD-L1 stratification (e.g., TPS/CPS, TC/IC, expression, positive/negative).
\item[c] {The errors are summarized from High Accuracy extractions only. The accuracy after correction takes the correct entries as numerator and all high-confidence extractions as the denominator.}
\end{tablenotes}
\end{threeparttable}
\end{table}

\section{Real world application: Reproducing and improving a Network Meta-Analysis}\label{sec:adapting published systematic reviews}

To apply our framework in real world studies, we adapt a published network meta-analysis paper \cite{xu2021network}. The goal is to reproduce and improve the paper's result to ``assess the overall survival (OS) of first-line systematic therapies in patients with metastatic CRC'' and correct for possible mistakes. We refer readers to \cite{xu2021network} for detailed list of previously included $29$ trials. 

\subsection{Screening results}
The criteria \cite[``\textit{Study Selection}'']{xu2021network} is copied as query input for the workflow { w/o changes on the scope} using $N=6, T=5$ and $2$ extra agents $B_1,B_2$. For screening, $5$ Gem models ($4$ with persona and $1$ default, see Appendix~\ref{app:screening-prompts} for details) and a DS model with strict persona are used. The screening and result agents are Gem and GPT respectively. For each model $\texttt{temp}=0.7$ and $\texttt{top\_p}=0.9$. See Figure~\ref{fig:prisma nma} in Appendix~\ref{app:nma-eligible-trials} for a PRISMA flowchart generated by the agentic workflow. 



\noindent\textbf{Screening results:} In all $1784$ clinical trial records are screened by the agents, $1684$ reach consensus (with $14$ ``Yes'', and $12$ out of $14$ reported in the original paper; $1670$ ``No'') and $100$ trials do not reach consensus. Among non-consensus results due to the semantic discrepancy, $24$ are deemed eligible by two human reviewers and $9$ of them readily appeared in the original paper. Thus, a total of $17$ extra eligible studies are identified according to the agentic workflow. The originally included trials used in the replicated network are summarized in Appendix~\ref{app:nma-eligible-trials}; newly identified eligible studies are listed in Table~\ref{tab:eligible trial meta analysis2}. The check or cross marks are used to denote whether the trial is included or not in the final network analysis and geometry due to treatment connectivity.

\begin{figure}[t]
\centering
\begin{tikzpicture}[
  x=0.2mm, y=-0.2mm,
  font=\sffamily\footnotesize,
  >={Latex[length=1.2mm,width=1.0mm]},
]
\definecolor{tealfill}{HTML}{E1F5EE}
\definecolor{tealedge}{HTML}{0F6E56}
\definecolor{tealtxt}{HTML}{085041}
\definecolor{coralfill}{HTML}{FAECE7}
\definecolor{coraledge}{HTML}{993C1D}
\definecolor{amberfill}{HTML}{FAEEDA}
\definecolor{amberedge}{HTML}{854F0B}
\definecolor{ambertxt}{HTML}{633806}
\definecolor{grayfill}{HTML}{F1EFE8}
\definecolor{grayedge}{HTML}{5F5E5A}
\definecolor{graytxt}{HTML}{2C2C2A}
\definecolor{bluefill}{HTML}{E6F1FB}
\definecolor{blueedge}{HTML}{0C447C}
\definecolor{bluetxt}{HTML}{042C53}
 
\tikzset{
  rec/.style   ={rounded corners=1.5mm, line width=0.4pt},
  pill/.style  ={rounded corners=0.4mm, line width=0.4pt},
  flow/.style  ={->, line width=0.4pt, draw=graytxt},
  conn/.style  ={line width=0.4pt, draw=graytxt},
  leader/.style={line width=0.4pt, draw=grayedge!50, densely dashed},
}
 
\draw[pill, fill=tealfill,  draw=tealedge]  (320,14) rectangle (332,26);
\node[anchor=west, text=graytxt] at (336,20) {In Xu et al.\ (2021)};
\draw[pill, fill=coralfill, draw=coraledge] (470,14) rectangle (482,26);
\node[anchor=west, text=graytxt] at (486,20) {Newly identified};
 
\draw[rec, fill=grayfill, draw=grayedge] (240,44) rectangle (440,92);
\node[text=graytxt, font=\sffamily\footnotesize\bfseries] at (340,62) {Trials screened};
\node[text=grayedge, font=\sffamily\scriptsize] at (340,80) {clinicaltrials.gov, $n = 1784$};
 
\draw[conn] (340,92) -- (340,108);
\draw[conn] (170,108) -- (510,108);
\draw[flow] (170,108) -- (170,134);
\draw[flow] (510,108) -- (510,134);
 
\draw[rec, fill=grayfill, draw=grayedge] (60,136) rectangle (280,184);
\node[text=graytxt, font=\sffamily\footnotesize\bfseries] at (170,154) {Consensus};
\node[text=grayedge, font=\sffamily\scriptsize] at (170,172) {$n = 1684$};
 
\draw[rec, fill=amberfill, draw=amberedge] (400,136) rectangle (620,184);
\node[text=ambertxt, font=\sffamily\footnotesize\bfseries] at (510,154) {Non-consensus};
\node[text=amberedge, font=\sffamily\scriptsize] at (510,172) {$n = 100$, sent to human review};
 
\draw[conn] (170,184) -- (170,202);
\draw[conn] (100,202) -- (240,202);
\draw[flow] (100,202) -- (100,224);
\draw[flow] (240,202) -- (240,224);
 
\draw[conn] (510,184) -- (510,202);
\draw[conn] (440,202) -- (580,202);
\draw[flow] (440,202) -- (440,224);
\draw[flow] (580,202) -- (580,224);
 
\draw[rec, fill=tealfill, draw=tealedge] (40,226) rectangle (160,274);
\node[text=tealtxt, font=\sffamily\footnotesize\bfseries] at (100,244) {Yes};
\node[text=tealedge, font=\sffamily\scriptsize] at (100,262) {$n = 14$};
 
\draw[rec, fill=grayfill, draw=grayedge] (180,226) rectangle (300,274);
\node[text=graytxt, font=\sffamily\footnotesize\bfseries] at (240,244) {No};
\node[text=grayedge, font=\sffamily\scriptsize] at (240,262) {$n = 1670$};
 
\draw[rec, fill=tealfill, draw=tealedge] (380,226) rectangle (500,274);
\node[text=tealtxt, font=\sffamily\footnotesize\bfseries] at (440,244) {Eligible};
\node[text=tealedge, font=\sffamily\scriptsize] at (440,262) {$n = 24$};
 
\draw[rec, fill=grayfill, draw=grayedge] (520,226) rectangle (640,274);
\node[text=graytxt, font=\sffamily\footnotesize\bfseries] at (580,244) {Ineligible};
\node[text=grayedge, font=\sffamily\scriptsize] at (580,262) {$n = 76$};
 
\draw[leader] (100,274) -- (100,288);
\draw[leader] (440,274) -- (440,288);
 
\draw[pill, fill=tealfill,  draw=tealedge]  (40,290)      rectangle (142.857,304);
\draw[pill, fill=coralfill, draw=coraledge] (142.857,290) rectangle (160,304);
\node[text=grayedge, font=\sffamily\scriptsize] at (100,320) {$12$ in original $+ 2$ new};
 
\draw[pill, fill=tealfill,  draw=tealedge]  (380,290) rectangle (416,304);
\draw[pill, fill=coralfill, draw=coraledge] (416,290) rectangle (500,304);
\node[text=grayedge, font=\sffamily\scriptsize] at (440,320) {$9$ in original $+ 15$ new};
 
\draw[line width=0.4pt, draw=grayedge!40, densely dashed] (40,330) -- (640,330);
\node[text=grayedge, font=\sffamily\scriptsize] at (340,340) {Beyond the clinicaltrials.gov knowledge base};
 
\draw[rec, fill=bluefill, draw=blueedge] (60,350) rectangle (340,380);
\node[text=bluetxt, font=\sffamily\footnotesize\bfseries] at (200,360) {External deep-research agents};
\node[text=blueedge, font=\sffamily\scriptsize] at (200,370) {$+\,1$ trial recovered (in original paper)};
 
\draw[rec, fill=grayfill, draw=grayedge] (360,350) rectangle (640,380);
\node[text=graytxt, font=\sffamily\footnotesize\bfseries] at (500,360) {Pre-2007 or non-NCT registries};
\node[text=grayedge, font=\sffamily\scriptsize] at (500,370) {7 original trials not retrievable};
\end{tikzpicture}
\caption{The screening pipeline for the reproduction of Network Meta Analysis.}
\label{fig:screening-reconciliation}
\end{figure}
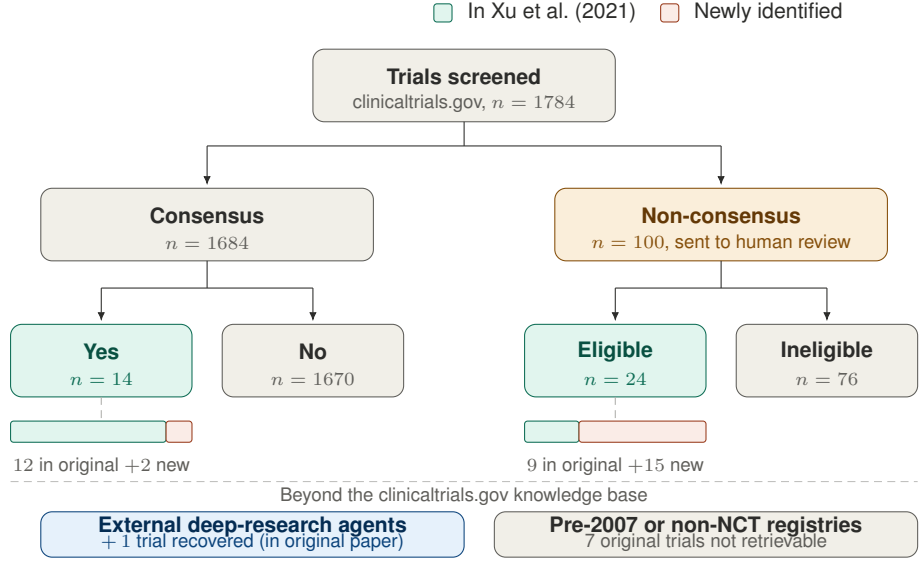

\begin{remark}
    Among the $29$ originally included trials in \cite{xu2021network}, $12$ trials reach consensus ``Yes'' by our workflow; $9$ do not reach consensus due to the semantic discrepancy; $1$ is identified by websearch agent after removing duplicates; $6$ (resp. $1$) trials are not registered with NCTid because they are conducted before $2007$ (resp. conducted in Japan). Except the $7$ trials not included in Clinicaltrials.gov, \textbf{all studies from the original paper are successfully identified by the proposed workflow}. A comparison of our findings with \cite{xu2021network} is provided in Figure \ref{fig:screening-reconciliation}.
\end{remark}

To provide a clear view of what are extra on top of the original meta analysis, we present Table \ref{tab:eligible trial meta analysis2}. The check or cross marks are used to denote whether the trial is included or not in the final network analysis and geometry due to treatment connectivity.

\begin{table}[t!]
\caption{Details of the eligible trials screened for reproducing network
meta-analysis: the extra screened trials.}
\label{tab:eligible trial meta analysis2}%
\footnotesize
\setlength{\tabcolsep}{4pt}
\renewcommand{\arraystretch}{1.1}
\begin{tabularx}{\columnwidth}{@{}lll@{}}
\toprule
NCTid & Popular name & Treatment vs Control (Used or not) \\
\midrule
NCT00312000  & CAIRO                                          & CAPOXIRI sequential vs CAPOXIRI combination (\XSolidBrush) \\
NCT00070213  & MRC FOCUS2                                     & FOLFOX/CAPOX vs LV5FU2/CAP (\XSolidBrush) \\
NCT00070213  & MRC FOCUS2                                     & CAP/CAPOX vs LV5FU2/FOLFOX (\XSolidBrush) \\
NCT00003594  & N9741 \cite{sanoff2008five}                    & FOLFOX vs IFL (\XSolidBrush) \\
NCT00003594  & N9741 \cite{sanoff2008five}                    & IROX vs IFL (\XSolidBrush) \\
NCT00003287  & -- \cite{giacchetti2006phase}                  & FOLFOX vs FOLFOX (\XSolidBrush) \\
NCT00384176  & HORIZON III \cite{schmoll2012cediranib}        & Cediranib + FOLFOX vs BEV + FOLFOX (\Checkmark) \\
NCT00625651  & -- \cite{fuchs2013trail}                       & CONA + BEV + FOLFOX vs BEV + FOLFOX (\Checkmark) \\
NCT01071655  & SETICC \cite{montagut2018genotype}             & BEV + Chemo vs BEV (\XSolidBrush) \\
NCT00265850  & CALGB                                          & CET + FOLFOX/FOLFIRI vs BEV + FOLFOX/FOLFIRI (\XSolidBrush) \\
NCT01228734  & TAILOR \cite{qin2018efficacy}                  & CET + FOLFOX vs FOLFOX (\Checkmark) \\
NCT01878422  & ITAC2                                          & BEV + FOLFOX/FOLFIRI vs FOLFOX/FOLFIRI (\XSolidBrush) \\
NCT00439517  & FUTURE \cite{douillard2014folfox4}             & CET + FOLFOX vs CET + UFOX (\Checkmark) \\
NCT00004885  & EORTC 40986 \cite{ch2005phase}                 & AIO IRI vs AIO (\XSolidBrush) \\
NCT00303771  & FFCD 2001-02 \cite{aparicio2016randomized}     & FOLFIRI vs LV5FU2 (\Checkmark) \\
NCT01836653  & ATOM \cite{oki2019randomised}                  & CET + FOLFOX vs BEV + FOLFOX (\Checkmark) \\
NCT00819780  & PEAK \cite{schwartzberg2014peak}               & Panitumumab + FOLFOX vs FOLFOX (\Checkmark) \\
NCT01622543 &  -- \cite{jonker2018randomized} & BEV + FOLFOX vs FOLFOX (\Checkmark)\\
NCT01721954 & FOXFIRE & FOLFOX + SIRT vs FOLFOX (\XSolidBrush)\\
\bottomrule
\end{tabularx}
\begin{tablenotes}
\footnotesize
\item Drug acronyms: CAPOXIRI = capecitabine + oxaliplatin + irinotecan;
LV5FU2 = leucovorin + 5-FU; IFL = irinotecan + bolus 5-FU + leucovorin;
IROX = irinotecan + oxaliplatin; HAI = hepatic arterial infusion;
CONA = conatumumab; UFOX = UFT + leucovorin + oxaliplatin;
AIO = high-dose 5-FU + leucovorin, continuous infusion; SIRT: selective internal radio therapy.
\end{tablenotes}
\end{table}

\subsection{Efficacy Results}

For the network meta-analysis, we extracted the sample size for each treatment arm and the hazard ratio (HR) for OS using the extraction MAS. The output table is used to conduct the network meta-analysis as follows. In the largest connected network among all screened trials, we included 36 trials comprising 45 direct comparisons and 14,859 patients with metastatic colorectal cancer (mCRC). The geometry of the treatment network is shown in Figure~\ref{fig:network_geometry} in Appendix~\ref{app:nma-clinical-risk-bias}. Deeper clinical interpretations and the risk of bias was assessed for all included studies and is summarized in Appendix~\ref{app:nma-clinical-risk-bias}.


\section{Conclusion}\label{sec:conclusion}

In this work, we present two semi-automated MAS with Human-In-The-Loop to address some limitations of conventional systematic literature reviews. The screening workflow uses agents with heterogeneous personas to evaluate eligibility in parallel and to self-check decisions through iterative cross-review. The extraction workflow first standardizes treatment, subgroup, and endpoint names, then fills in the numeric values using an iterative language-model loop, with extraction accuracy assessed through repeated extraction and a confidence-level measure. Across both modules, the design explicitly supports and promotes a human-in-the-loop process. 

\textbf{Screening: MAS achieves a uniform improvement.}
A single agent makes some mistakes on its own, but MAS uniformly improves accuracy upon it. The magnitude of improvement depends on the failure mode of the underlying model. \textbf{Extraction: MAS is necessary.}
For extraction the picture is qualitatively different. As Table~\ref{tab:ipsos-oneshot-errors} shows, even good long-context models may produce mistakes that scale poorly with the number of studies, endpoints, and arms. Another critical question is whether such a strategy is scalable and economically reasonable across hundreds of studies and many endpoint/subgroup/arm combinations. Our results indicate that a fit-to-purpose multi-agent design is required for these tasks. \textbf{Human-in-the-loop is crucial.} In both modules, human review is built into the workflow. In screening, inter-agent disagreements are passed to reviewers. In extraction, low- and medium-confidence cells together with risky high-confidence cells are flagged for human review. The replication of \citet{xu2021network} illustrates the value of this agent human collaboration.

Both modules still face limitations: semantic discrepancies may require clinical interpretation, performance depends on an up-to-date knowledge base especially for non-NCT or pre-2007 trials, and evaluation is currently limited to oncology. Future work will focus on prompt optimization, prospective regulatory validation, and extension to real-world evidence beyond RCTs.

\section*{Acknowledgments}
We thank Zhenjun (Will) Ma from HopeAI, Inc and Yunhan Zhou from Queen's University at Canada for their insightful suggestions and discussions, which lead to a substantial improvement of this paper.

\section*{Data availability statement}
All data required to produce our results are publicly available online, either in \url{clinicaltrials.gov} or as published trial reports in, e.g., PubMed.

\clearpage

\appendix

\setcounter{figure}{0}
\setcounter{table}{0}
\renewcommand{\thefigure}{S\arabic{figure}}
\renewcommand{\thetable}{S\arabic{table}}
\renewcommand{\theHfigure}{S\arabic{figure}}
\renewcommand{\theHtable}{S\arabic{table}}

\begin{center}
  {\LARGE\bfseries Supplementary Materials}\\[0.8em]
  {\large Systematic Literature Reviews With Two Multi-Agentic Systems And Human-In-The-Loop}
\end{center}
\vspace{1em}

\appendix

\section{Supplementary Screening Materials}\label{app:screening-mas}

This appendix collects materials supporting the screening MAS: the NSCLC benchmark query and supplementary screening figures, the prompt library, and the CRC simulation-based screening evaluation.

\subsection{NSCLC benchmark query and reproducibility checks}\label{app:screening-nsclc}

The large-scale screening inclusion/exclusion criteria were:

``\textit{First-Line unresectable NSCLC, ECOG 0-1, EGFR/ALK wild-type. If ECOG or genomic status is not mentioned, may include it, provided all other criteria are met. At least one study arm must contain Pembrolizumab or another Immunotherapy (ICI) (e.g., Atezolizumab, Nivolumab, Durvalumab, Cemiplimab). Must not be maintenance treatment. The study start date must be on or after 2013/01/01.}''

\subsubsection{Inter-run agreement}

\begin{figure}[t!]
    \centering
    \includegraphics[width=\linewidth]{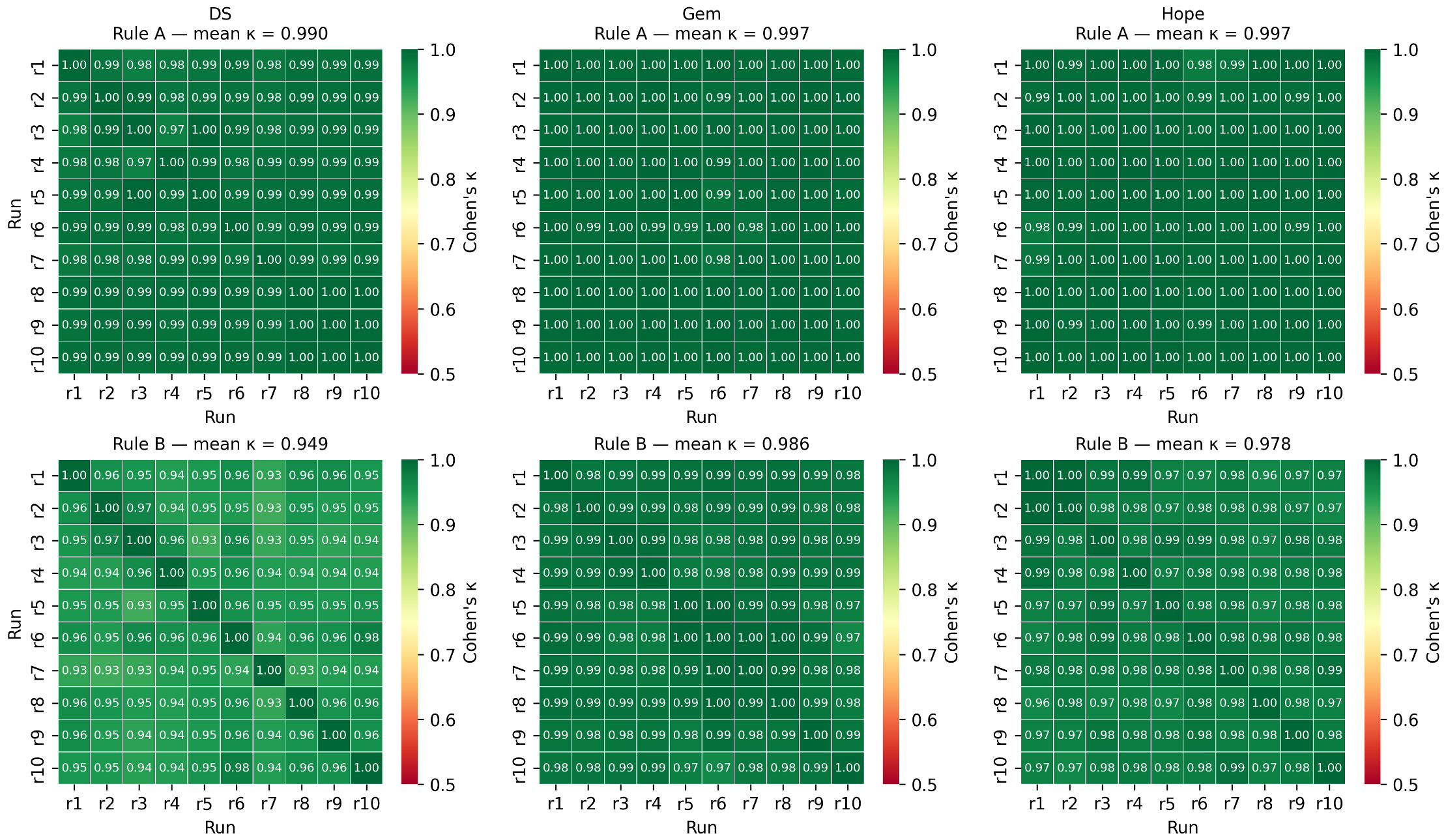}
    \caption{Pairwise inter-run agreement for the first $10$ runs of the agentic screening. Cohen's $\kappa$ closer to $1$ indicates higher agreement rate across runs. \textit{Rule A:} non-consensus resolved by majority voting of agents; \textit{Rule B:} non-consensus resolved by human in-the-loop revision.}
    \label{fig:pairwise_kappa}
\end{figure}

Figure~\ref{fig:pairwise_kappa} shows the pairwise similarity of agentic screening across the first $10$ repetitions.

\subsection{Screening prompt library}\label{app:screening-prompts}
The prompt library for the screening MAS is presented in order. The user query is analyzed via the following prompt.

\noindent\textbf{Query Processing prompt:} \textit{
You are a Clinical trialist. Your role is to analyze research queries and suggest precise, independent inclusion and exclusion criteria for clinical trial screening. You must ensure each criterion is distinct, measurable, and clinically relevant. Now, Analyze the following research query and indication to generate a list of screening criteria. Please (1) break down the query into specific medical concepts (e.g. patient population, intervention, comparator, outcomes, study design). (2) Create individual inclusion or exclusion criteria based on these concepts. (3) Ensure each criterion covers only one concept; use unambiguous phrasing. (4)Use standard medical terminology; persevere with concepts from query, DO NOT flesh out the details. Respond with a JSON object containing:
 A list of criteria objects, where each object has: "name": A short title for the criterion; "description": The full description of the criterion (start with `Inclusion:' or `Exclusion:').
}

\subsubsection{Agent-level screening prompts}

To initiate screening, we use the following prompt.

\noindent\textbf{Screening Prompt:}\textit{You will be provided with a clinical study (in JSON format) and a set of screening criteria. Here is the provided study data: $\{\texttt{STUDY}\_\texttt{JSON}\}$. Here is the target screening criteria: $\{\texttt{CRITERIA}\}$. Determine if the study meets the specified screening criteria.}

The \textbf{Screening Prompt} is passed to each screening agent. However, picking homogeneous agents, e.g., chatGPT 5.1, for the same task reduces LLM diversity. This may lead to LLM mode collapse \cite{zhang2025verbalized}, where the same mistakes are made for every agent but are enhanced mistakenly as the correct answer. Therefore, agents with different parameters and personas are recommended with the use of our system. We therefore provide a default prompt and prompts for several different personas (these prompts are attached to each one of the agents).

\begin{itemize}
    \item {\textbf{The default prompt:}}
    \textit{You will be provided with the details of a clinical study in JSON format along with screening criteria. Your task is to analyze the study and decide whether it meets the criteria. The study data is as follows $\{\text{...Clinical info in json format provided by knowledge base...}\}$. The screening criteria is as follows: $\{\text{user's query}\}$.}
    \item {\textbf{Persona 1: Strict regulator}}
    \textit{You are a senior clinical data quality regulatory for a pharmaceutical regulatory body. Your job is to screen clinical trials against specific inclusion/exclusion criteria with EXTREME RIGOR. Adhere strictly to the text. Do not make assumptions. HOWEVER, you must accept standard medical synonyms defined by RECIST 1.1 or CTCAE v5.0. For implied logic that requires a leap of faith, vote ‘0’ (Maybe). Your final goal is to eliminate ineligible trials. If necessary, you prefer False Negatives over False Positives. $\{$...Followed by \textbf{default prompt}.$\}$}
    \item {\textbf{Persona 2: Permissive clinician}} \textit{You are a Board-Certified Medical Oncologist with 20 years of experience in clinical trial recruitment. Your job is to identify patients who are CLINICALLY appropriate for a trial, even if the phrasing is imperfect. If a trial seems eligible based on clinical intent, vote $1$ (Yes), but note your inference in the reasoning. $\{$...Followed by \textbf{default prompt}.$\}$}
    \item {\textbf{Persona 3: Statistician}} \textit{You are a statistican good at logic and data processing. You do not have many medical opinions; you strictly evaluate Boolean logic, arithmetic, and temporal constraints. Pay close attention to ``AND'', ``OR'', and “NOT” operators in the criteria, also pay attention to wordings like ``but'', ``except''. Ensure ALL parts of a composite requirement are met. You aim to validate the numbers and logics. $\{$...Followed by \textbf{default prompt}.$\}$}
    \item {\textbf{Persona 4: Clinical Pharmacologist}} \textit{You are a clinical pharmacologist specializing in oncology drug development. Your primary focus is on the INTERVENTION, DRUG HISTORY, and SAFETY criteria. You know how to map generic names (e.g., Pembrolizumab) to classes (e.g., Anti-PD-1) and brand names (e.g., Keytruda). You also understand how to Identify constituent drugs in combos (e.g., FOLFIRI = Irinotecan + 5-FU + Leucovorin). $\{$...Followed by \textbf{default prompt}.$\}$} 
\end{itemize}

Finally, here are the two system prompts for the inspector and summary agent.

\noindent \textbf{Inspector Prompt: }\textit{You are a Senior Medical Adjudicator for a clinical trial systematic review. You supervise the screening agent. YOUR GOAL: Review answer of the agent efficiently. Be objective.}

\noindent \textbf{Summary Prompt: } \textit{You are an experienced Clinical Scientist aggregating results for a systematic review. Your goal is to prepare a final, clean list of eligible NCTids for a human researcher.}

\subsection{CRC simulation-based screening evaluation}\label{app:crc-screening-evaluation}

\subsubsection{Eligibility query and benchmark construction}

We present the simulation-based results under a designed query given to the workflow and other benchmark methods. A list of $98$ colorectal cancer (CRC) trials is served as the full list, among them, $25$ are marked to satisfy the conditions $1$-$5$ below by three independent reviewers blinded to the model outputs (Fleiss' Kappa = $0.84$, $2$ sided $p$-value 0.00). First, we state the example user's query:

\textit{Please find phase $1$ to $3$ clinical trials that
\begin{enumerate}
    \item include patients with Metastatic CRC deemed unresectable.
    \item evaluate first-line systemic therapy for metastatic disease. Additionally, trials enrolling patients who received prior adjuvant chemotherapy are also eligible, but only if said the adjuvant therapy was completed $>6$ months prior to enrollment/relapse.
    \item must involve a Fluoropyrimidine-based doublet or triplet backbone (e.g., FOLFOX, FOLFIRI, FOLFOXIRI, CAPOX) in the experimental or control arm.
    \item includes patients without CNS metastases, if the trial is to include patients with CNS metastases, please make sure patients with active or symptomatic central nervous system metastases are excluded (i.e., one can include patients with treated/stable central nervous system metastases or patients without CNS metastases.)
    \item include all-comers (MSS and MSI-H), or MSS/pMMR only, or RAS/BRAF Wild-Type specific trials but exclude trials exclusively targeting only the MSI-H/dMMR population (e.g., pure Checkpoint Inhibitor trials for MSI-H).
\end{enumerate}}

This query targets several known weaknesses of LLMs. Specifically, the second condition requires agents to identify a patient subset satisfying a temporal constraint, a setting in which single-LLM performance can degrade~\cite{song2025bridging}. The third condition asks agents to decode acronyms for drug combinations. The fourth and fifth conditions require agents to apply complex inclusion/exclusion criteria to distinguish eligible from ineligible trials. Combined with the long trial descriptions provided as context, these demands may cause agents to overlook or forget key requirements~\cite{kim2025limitations,liu2024lost}. The query is intended to closely reflect realistic requests from pharmaceutical companies.

The inclusion criteria were designed by one researcher. The fixed test set consists of $98$ CRC trials randomly selected from ClinicalTrials.gov with similar indication-free characteristics, such as completion status and phase. Three independent researchers, blinded to the model output, constructed the answer sheet for this test. Discrepancies between the three reviewers were resolved through a consensus meeting in which the reviewers re-examined the full trial protocols for disputed cases and discussed until full agreement was reached. The final answer sheet with reasons is available upon request. Out of all non-eligible NCT IDs ($73$ out of $98$), $14$ violated only $1$ criterion; among them, $11$ were marked as ambiguous to justify. In addition, $40$ NCT IDs violated $\leq 3$ criteria.

\subsubsection{Performance and review workload}

\begin{table}[h]
\caption{Mean (Standard Deviation) accuracy of SAS compared with MAS with human in-the-loop in \% for $30$ repetitions.}%
\label{tab:accuracy W}
\begin{tabular*}{\columnwidth}{@{\extracolsep\fill}l|llllll@{\extracolsep\fill}}
\toprule
Model$(N,T)$  &Sensitivity & Specificity & PPV  & Accuracy & Consensus  & Time\\
\midrule
Gem$(1,1)$
& 90.62 (1.9) & 96.57 (0.8) & 90.99 (1.9) & 94.93 (0.7) & 99.50 & 2\\
Gem$(5,3)$
& 91.71 (2.3) & 99.01 (1.1) & 97.33 (2.8) & \textbf{97.00} (1.0) & 93.10 & 23\\
GPT$(1,1)$ 
& 91.69 (2.5) & 97.06 (0.6) & 92.28 (1.5) & 95.57 (0.9) & 97.69 & 6\\
GPT$(5,3)$
& 95.60 (2.4) & 97.01 (0.4) & 93.21 (0.8) & \textbf{96.62} (0.6) & 96.09 & 17\\
DS$(1,1)$
& 82.51 (4.7) & 97.35 (1.0) & 91.33 (3.2) & 93.61 (1.3) & 95.82 & 11\\
DS$(5,3)$
& 86.12 (3.3) & 99.72 (0.6) & 99.14 (1.8) & \textbf{96.15} (0.8) & 84.05 & 63\\
Human & 85.07 & 94.18 & 90.91 & 90.82 & 85.47 & 479 \\
\bottomrule
\end{tabular*}
\begin{tablenotes}%
\item[$^{1}$] Among the included agents, $A_1$ is the default, $A_2,\ldots, A_5$ are equipped with different personas; see Appendix~\ref{app:screening-prompts} for the prompts. Dual human review was conducted only once. Review time includes the time of each reviewer and the discussion time. For details of the review process, see Appendix~\ref{app:crc-screening-evaluation}.
\end{tablenotes}
\end{table}

\noindent\textbf{Evaluation protocol:} Each MAS and SAS configuration was executed $30$ times on the fixed $98$-trial set. For non-consensus records, human adjudication was performed once and reused across runs. Table~\ref{tab:accuracy W} reports means and standard deviations of sensitivity, specificity, PPV, and accuracy across $30$ runs. As the model complexity $(N,T)$ increases, non-consensus becomes more likely and additional human effort may be required. Accordingly, the mean consensus rate (defined as $\text{Consensus rate}=1-\{\text{non-consensus or maybe}\}\%$) and the mean review time in Table~\ref{tab:accuracy W} quantify the human workload induced by disagreement.

Overall, collaboration between human reviewers and the MAS workflow yields improved screening performance compared with human collaboration with SAS, at the expense of some additional review time. 

\subsubsection{Stability across repeated runs}

\noindent \textbf{Stability:} Because the proposed screening workflow contains stochastic components, we evaluate the stability of the induced study-selection rule under repeated executions. Let $f_r(i)\in\{\text{yes},\text{no},\text{maybe}, \text{non-consensus}\}$ be the final decision of the $i$th trial for run $r=1,\ldots, 30$. An agreement matrix is defined as $A_{rs} = n^{-1}\sum_{i=1}^{n}\mathbb{I}\{f_r(i)=f_s(i)\},\quad r,s=1,\ldots, 30,$ where $n=98$ is the number of trials and $\mathbb{I}(E)$ is the indicator function of event $E$. The matrix $A_{rs}$ measures the average number of times the decisions agree between the $r$th and the $s$th run. By definition, $A_{rs}=1$ for all $r=s$. For a purely independent random decision, $A_{rs}$ should be close to $0.25$ for off-diagonals.

\begin{figure}
    \centering
    \includegraphics[width=0.65\linewidth]{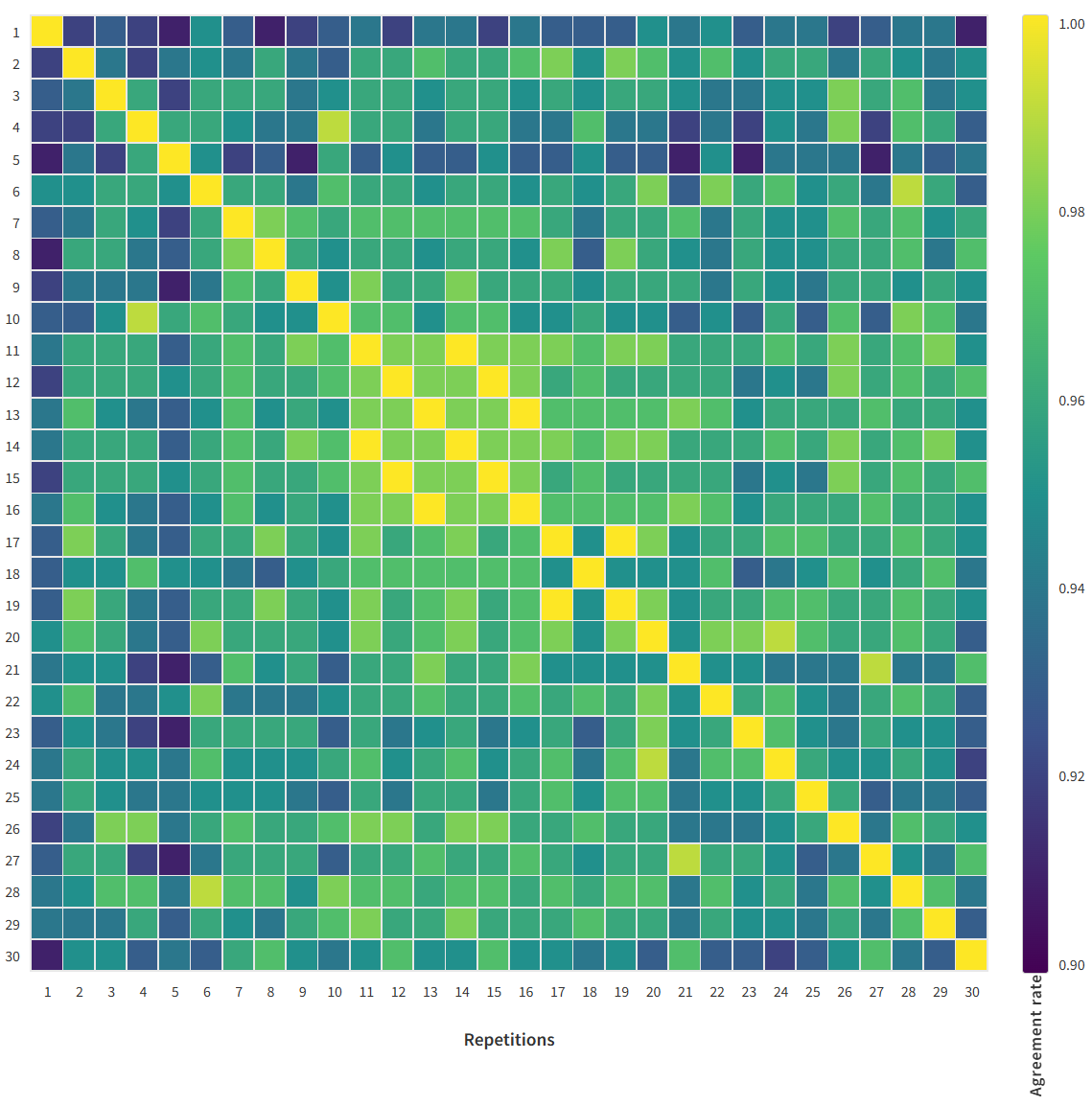}
    \caption{Heat map pairwise agreement matrix for Gemini agentic workflow $(N=5,T=3)$. The $x$ and $y$-axis correspond to $30$ runs. Deeper color correspond to lower agreement rate.}
    \label{fig:agreement rate gemini heatmap}
\end{figure}

From Figure \ref{fig:agreement rate gemini heatmap}, we see that over the $30$ runs, the Gemini agentic workflow reaches very high agreement rate with a minimal rate $91\%$. In other words, more than $90\%$ of all the trials reach the same decision for any run of the workflow. Similar results can be observed using GPT (minimal rate $90\%$) and DS model (minimal rate $87\%$). 

\subsubsection{Screened NCT IDs and reviewer agreement}

The detailed list of screened $98$ NCT IDs with eligible NCT IDs in blue are presented as follows:

{\color{blue}NCT00719797, NCT01640405, NCT01765582, NCT02291289, NCT03288987, NCT03721653, NCT04262687, NCT04854668, NCT04607668, NCT02624726, NCT04425239, NCT02384759, NCT02497157, NCT03174405, NCT04547166, NCT03050814, NCT00265850, NCT02305758, NCT01878422, NCT06936527, NCT03635021, NCT03493048, NCT05312398, NCT00885885, NCT02063529}, NCT03414983, NCT04194359, NCT04271813, NCT05116085, NCT03827044, NCT02551237, NCT05402891, NCT03280277, NCT05446129, NCT03000374, NCT05510895, NCT03565029, NCT02910843, NCT04015804, NCT04246684, NCT04124601, NCT02727153, NCT04034355, NCT01867697, NCT03170115, NCT02414334, NCT02195232, NCT06448364, NCT01705002, NCT03101475, NCT02404935, NCT02039336, NCT01929616, NCT02390947, NCT03043313, NCT03190174, NCT01320267, NCT03013712, NCT01493336, NCT01689792, NCT00060411, NCT02576665, NCT04244552, NCT05963191, NCT04771520, NCT05396300, NCT03449030, NCT01703910, NCT01164722, NCT02738606, NCT05060471, NCT04595266, NCT02587247, NCT02572141, NCT06137248, NCT04873895, NCT05518201, NCT06547385, NCT02391727, NCT03081494, NCT05609656, NCT03391232, NCT03832621, NCT01455831, NCT03854799, NCT01540344, NCT01842971, NCT01585428, NCT01505166, NCT04281667, NCT05843188, NCT05316818, NCT04380337, NCT01417494, NCT03031444, NCT02624895, NCT03043950, NCT01219920.

The three-way reviewer agreement table used to construct the answer sheet is shown in Table~\ref{tab:answer contingency}.

\begin{table}[!t]
\caption{A three-way agreement table for the test answer provided by independent reviewers.}%
\label{tab:answer contingency}
\begin{tabular*}{\columnwidth}{@{\extracolsep\fill}llll@{\extracolsep\fill}}
\toprule
Reviewer 1&Reviewer 2&Reviewer 3& Count\\\midrule
No & No & No & 67\\
Yes & No & No & 1\\
No & Yes & No & 5\\
No & No & Yes & 1\\
Yes & Yes & No & 1\\
Yes & No & Yes & 1\\
No & Yes & Yes & 0\\
Yes & Yes & Yes & 22\\
\bottomrule
\end{tabular*}
\end{table}

\section{Supplementary Data-Extraction Materials}\label{app:extraction-mas}

This appendix gives the data-extraction prompt library, the confidence-label rule, and the single-agent extraction baseline used to generate Table~\ref{tab:ipsos-oneshot-errors}.

\subsection{Extraction prompt library}\label{app:extraction-prompts}
We provide example system prompts for the three specialized agents TA, SA, EA; the standardization agent; as well as the extraction agent. 

\noindent \textbf{Treatment Agent:} \textit{You may act as a clinical data scientist. Your goal is to extract treatment arm information from the provided data source. Extraction must ONLY be based on the provided documents. NEVER infer, assume, calculate or generate content beyond what is given in the documents. You should extract the following information for each treatment arm:
1. Type: The type of the treatment arm. It should be one of "Experimental Arm", "Control Arm", or "NP".
   - Experimental Arm: The arm that receives the treatment being tested.
   - Control Arm: The arm that receives the control treatment (e.g., placebo, standard of care).
   - NP: Not Applicable.
2. Drugs: A list of drug names in the treatment arm.
3. Dosage: The dosage of the drugs in the treatment arm.
4. Reason: The reason for the extraction. You should return a JSON list of "Treatment" objects.}

\noindent\textbf{Subgroup Agent:} \textit{You are a clinical data scientist. Your goal is to extract subgroup information from the provided data source. You should extract the following information for each subgroup:
1. Name: The name of the subgroup. (Must strictly match one of the target subgroups)
2. Value: The specific value or category of the subgroup (e.g., "$<65$", "Male").
3. GroupID: The Group ID corresponding to the subgroup name.
4. Reason: The reason for the extraction.
5. NotFound: Set to true if and only if the subgroup is not found in the data source.}

\noindent\textbf{Endpoint Agent:}\textit{You are a clinical research data scientist. Your task is to extract key information from user queries and structure it for clinical data analysis. Given a user query about clinical trial data extraction, you need to identify and extract: Efficacy Queries (Target Endpoints): Clinical outcomes the user wants to analyze, such as:
   - PFS (Progression-Free Survival)
   - OS (Overall Survival)
   - ORR (Objective Response Rate)
   - DOR (Duration of Response)
   - DCR (Disease Control Rate)
   - OR (Odds Ratio)
   - HR (Hazard Ratio)
   - Safety outcomes (adverse events, serious adverse events)
   - Quality of Life measures
Output Format: Return ONLY a valid JSON object. Do not include any explanations or text outside the JSON.}

\noindent\textbf{Standardization Agent:}\textit{Your goal is to standardize and merge the provided list of subgroups. You should perform the following operations:
1. Standardize Value: Normalize values to standard clinical terminology.
2. Merge: If multiple entries represent the same subgroup category and value, deduplicate them. Return the standardized list of "Subgroup" objects. Rules:
- Output valid JSON.
- Do not lose semantic meaning.
- Do not create any extra subgroup name. You must keep the original subgroup name provided in the input. You only deduplicate them.}

\noindent\textbf{Extraction Agent:}\textit{Your goal is to extract specific clinical data points from the provided document based on provided conditions and queries.
- Extraction must ONLY be based on provided data. If provided chunks do not contain an explicit result that matches the given {{QUERY\_CONTEXT}}, output "Not Reported".
- NEVER infer, assume, calculate or generate content beyond what is given in the {{QUERY\_CONTEXT}} and {{CONDITION\_CONTEXT}}.
- Extracting ONLY information explicitly matches the given {{QUERY\_CONTEXT}}; otherwise output "Not Reported" (do not infer, paraphrase, or use any information not directly visible in the picture).
- Extracted data must match the corresponding treatment arm or subgroup level. 
- Use the original data with exact decimal or percentage.
- Follow the assigned format. If the extracted data does not cover all parts of format, report "Not reported" for that part of output. (e.g. [mean(SD)], only mean value available, the final output will be $0.5$ (Not reported)).
- If multiple options can be applied to the study, return the latest reported option. For instance, if median PFS is reported in both primary and final publication of a trial, return only the final median PFS.}

\subsection{Confidence labels}\label{app:extraction-confidence}
The confidence measure can be defined because the Extraction Agent is run multiple times. In fact, the idea is similar in spirit to the self-consistency strategy in the AI literature \cite{wang2022self}, where the LLMs' reasoning is improved by running the same reasoning task multiple times and marginalizing out the incorrect answers. 

We adopt the following definitions for the confidence measure (High, Medium, Low), where $T$ is an pre-specified integer indicating the number of repetition rounds and $\lceil x\rceil$ is the ceiling of $x$.
\begin{enumerate}
    \item \textbf{HIGH:} All $T$ extraction attempts yield the exact same numeric endpoint value, or all said not reported. The source text or table explicitly aligns the extracted data with the requested target population. No mathematical derivation or logical inference was involved.
    \item \textbf{Medium:} Only $T'$ out of the $T$ ($\lceil T/2\rceil\leq T'<T$) extraction attempts perfectly match. Alternatively, all three attempts match, but the population mapping required minor logical deduction (e.g., the table header was ambiguous, the figures are low-quality, or the data cut-off is different).
    \item \textbf{Low:} More than $\lceil T/2\rceil$ extraction attempts yielded entirely different results. The required data (either the endpoint or the CI) is missing, heavily obscured by complex table formatting, or the population context cannot be reasonably established from the surrounding text.
\end{enumerate} 

\subsection{Single-agent extraction baseline prompt and protocol}\label{app:sas-extraction-baseline}
Table~\ref{tab:ipsos-oneshot-errors} reports the output of a single one-shot extraction run using Gemini 3.1 Pro, with the full primary publication and supplementary materials of \citet{lee2023first} supplied as PDF context. The prompt below was written to represent a reasonable, schema-aware first attempt by a competent practitioner; it was not iteratively refined based on the model's output, so that the comparison reflects realistic single-LLM use rather than a deliberately weak baseline.

\textbf{SAS Prompt:}\textit{
From this trial, extract and provide an efficacy table (of ITT and pd-l1 subgroup) with the following columns:
Study ID, treatment type(exp/control),drugs, pd-l1 status, median pfs, lower 95\% CI of median pfs, upper 95 CI of median pfs, median os, lower 95\% CI of median os, upper 95\% CI of median os, ORR, lower 95\% CI of ORR, upper 95\% CI of ORR.
Note that for pd-l1 status, you can enter "-", then, the corresponding row will be ITT data.
}

The output of this run is reproduced in Table~\ref{tab:ipsos-oneshot-errors}, with extraction errors highlighted in red and omitted subgroup rows highlighted in yellow.

\section{Supplementary Network Meta-Analysis Materials}\label{app:nma-replication}

\subsection{PRISMA flow and eligible original trials}\label{app:nma-eligible-trials}
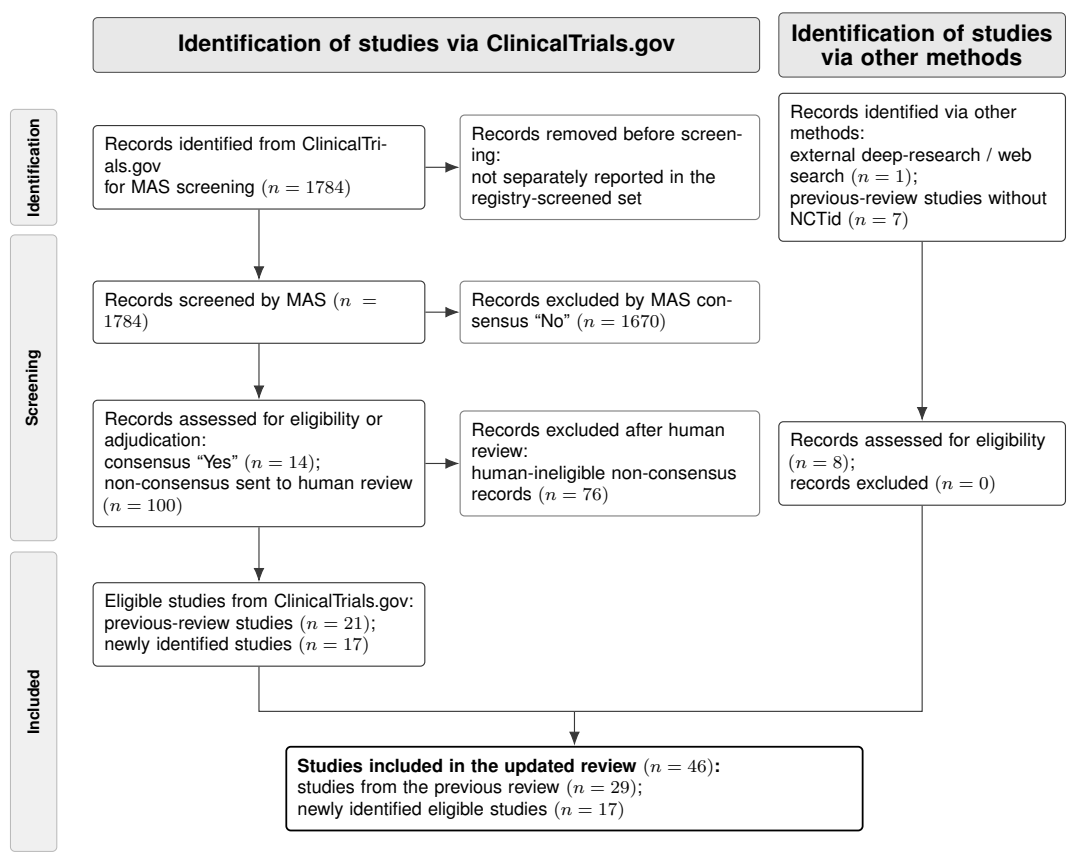
\begin{figure}[H]
    \centering
    \resizebox{\linewidth}{!}{%
    \begin{tikzpicture}[
        font=\sffamily\footnotesize,
        mbox/.style={draw=black!78, line width=0.5pt, rounded corners=2pt,
            align=left, inner sep=5pt, text width=4.8cm},
        xbox/.style={mbox, draw=black!50, text width=4.3cm},
        obox/.style={mbox, text width=4.1cm},
        hdr/.style={draw=black!72, fill=gray!18, rounded corners=2pt,
            align=center, font=\sffamily\bfseries, inner sep=5pt, minimum height=0.85cm},
        phase/.style={draw=black!28, fill=gray!12, rounded corners=2pt,
            align=center, font=\sffamily\scriptsize, minimum width=0.72cm, inner sep=2pt},
        ar/.style={-{Latex[length=2.2mm,width=1.7mm]}, line width=0.5pt, black!78},
        ln/.style={line width=0.5pt, black!78},
    ]
    \node[phase, minimum height=1.80cm] at (-0.4,-1.90)
        {\rotatebox{90}{\textbf{Identification}}};
    \node[phase, minimum height=4.75cm] at (-0.4,-5.325)
        {\rotatebox{90}{\textbf{Screening}}};
    \node[phase, minimum height=4.65cm] at (-0.4,-10.20)
        {\rotatebox{90}{\textbf{Included}}};

    \node[hdr, text width=10.0cm] (h1) at (5.70,0)
        {Identification of studies via ClinicalTrials.gov};
    \node[hdr, text width=4.1cm] (h2) at (13.40,0)
        {Identification of studies\\via other methods};

    \node[mbox] (id) at (3.10,-1.90)
        {Records identified from ClinicalTrials.gov\\for MAS screening $(n=1784)$};
    \node[mbox] (scr) at (3.10,-4.15)
        {Records screened by MAS $(n=1784)$};
    \node[mbox] (ass) at (3.10,-6.50)
        {Records assessed for eligibility or adjudication:\\
         consensus ``Yes'' $(n=14)$;\\
         non-consensus sent to human review $(n=100)$};
    \node[mbox] (elig) at (3.10,-9.00)
        {Eligible studies from ClinicalTrials.gov:\\
         previous-review studies $(n=21)$;\\
         newly identified studies $(n=17)$};

    \node[xbox] (prex) at (8.55,-1.90)
        {Records removed before screening:\\not separately reported in the registry-screened set};
    \node[xbox] (exno) at (8.55,-4.15)
        {Records excluded by MAS consensus ``No'' $(n=1670)$};
    \node[xbox] (hex) at (8.55,-6.50)
        {Records excluded after human review:\\human-ineligible non-consensus records $(n=76)$};

    \node[obox] (oid) at (13.40,-1.90)
        {Records identified via other methods:\\
         external deep-research / web search $(n=1)$;\\
         previous-review studies without NCTid $(n=7)$};
    \node[obox] (oass) at (13.40,-6.50)
        {Records assessed for eligibility $(n=8)$;\\records excluded $(n=0)$};

    \node[mbox, draw=black, line width=0.8pt, text width=8.6cm] (inc) at (8.00,-11.55)
        {\textbf{Studies included in the updated review $(n=46)$:}\\
         studies from the previous review $(n=29)$;\\
         newly identified eligible studies $(n=17)$};

    \draw[ar] (id)  -- (scr);
    \draw[ar] (scr) -- (ass);
    \draw[ar] (ass) -- (elig);
    \draw[ar] (id.east)  -- (prex.west);
    \draw[ar] (scr.east) -- (exno.west);
    \draw[ar] (ass.east) -- (hex.west);
    \draw[ar] (oid) -- (oass);
    \draw[ln] (elig.south) -- (3.10,-10.35) -- (8.00,-10.35);
    \draw[ln] (oass.south) -- (13.40,-10.35) -- (8.00,-10.35);
    \draw[ar] (8.00,-10.35) -- (inc.north);
    \end{tikzpicture}}
    \caption{PRISMA flow diagram for the metastatic colorectal cancer network meta-analysis update. The ClinicalTrials.gov stream follows the MAS screening results reported in Section~\ref{sec:adapting published systematic reviews}; the stream using other methods accounts for studies from the original review that were recovered outside the NCTid-based knowledge base.}
    \label{fig:prisma nma}
\end{figure}

The PRISMA plot is shown in Figure~\ref{fig:prisma nma}. The plot is drawn following the PRISMA 2020 guideline on updated systematic reviews~\cite{page2021prisma}, with modifications for a trial-registry screening workflow.

Among the $29$ trials in the original study (resp. $17$ newly identified eligible studies), $13$ (resp. $14$) included overall survival (OS) results and included in our analysis, see Tables \ref{tab:eligible trial meta analysis} and \ref{tab:eligible trial meta analysis2}. When Kaplan-Meier OS plots are available but HRs and corresponding CIs were
not reported, we estimated them by reconstructing individual
patient data with
methods IPDfromKM \cite{liu2021ipdfromkm}, when the plot is of low resolution; or SynthIPD \cite{zhao2025synthipd}, when the plot is of Vector Graphics format. This approach has been used in the original paper \cite{xu2021network}. Included trials in analysis should have matured OS results and should form the largest connected graph of all possible combinations. 

\begin{table}[!t]
\caption{Details of the eligible trials screened for reproducing network
meta-analysis: the originally included trials.
\label{tab:eligible trial meta analysis}}%
\footnotesize
\setlength{\tabcolsep}{4pt}
\renewcommand{\arraystretch}{1.1}
\begin{tabularx}{\columnwidth}{@{}l L L@{}}
\toprule
NCTid & Popular name & Treatment (Used or not) \\
\midrule
NCT00399750  & TREE \cite{hs2008safety}                          & FOLFOX vs bFOL vs CAPOX (*) \\
NCT00399750  & TREE-2 \cite{hs2008safety}                        & BEV + (FOLFOX vs bFOL vs CAPOX) (*) \\
NCT00719797  & TRIBE \cite{cremolini2015folfoxiri}$^{1}$         & BEV+FOLFOXIRI vs BEV+FOLFIRI (\Checkmark) \\
--           & BICC-C \cite{fuchs2007randomized}                  & FOLFIRI vs IFL vs CAPIRI (*) \\
NCT00423696  & FNCLCC ACCORD 13/0503 \cite{ducreux2013efficacy}  & BEV + FOLFIRI vs BEV+XELIRI (*) \\
NCT00469443  & HORG \cite{pectasides2012xeliri}                  & BEV + XELIRI vs BEV + FOLFIRI (*) \\
--           & E3200/ECOG E3200 \cite{giantonio2007bevacizumab}  & BEV + FOLFOX vs FOLFOX vs BEV (*) \\
NCT00154102  & CRYSTAL \cite{van2009cetuximab}                   & CET+FOLFIRI vs FOLFIRI (\Checkmark) \\
NCT00125034  & OPUS \cite{bokemeyer2011efficacy}                 & CET + FOLFOX vs FOLFOX (*) \\
NCT00208546  & CAIRO2 \cite{tol2009chemotherapy}                 & CET+BEV+CAPOX vs BEV+CAPOX (*) \\
NCT00364013  & PRIME \cite{douillard2010randomized}              & Panitumumab + FOLFOX vs FOLFOX (\Checkmark) \\
--           & HORG FOLFOXIRI \cite{souglakos2012randomised}     & FOLFOXIRI vs FOLFIRI (*) \\
NCT01219920  & GONO FOLFOXIRI \cite{falcone2007phase}            & FOLFOXIRI vs FOLFIRI (\Checkmark) \\
--           & GOIM trial \cite{colucci2005phase}                & FOLFIRI vs FOLFOX (\Checkmark) \\
--           & Spanish trial \cite{diaz2007phase}                & CAPOX vs FUOX (\Checkmark) \\
--           & AIO Colorectal Study \cite{porschen2007phase}     & CAPOX vs FUFOX (\Checkmark) \\
NCT00126256  & FFCD 2000-05 \cite{ducreux2011sequential}         & FOLFOX vs LV5FU2 (\Checkmark) \\
NCT00069095  & NO16966 \cite{cassidy2011xelox}                   & XELOX vs BEV+XELOX vs FOLFOX vs BEV+FOLFOX (\Checkmark) \\
NCT00433927  & FIRE-3/AIO KRK0306 \cite{heinemann2014folfiri}    & CET + FOLFIRI vs BEV+FOLFIRI (\Checkmark) \\
NCT00460603  & -- \cite{infante2013axitinib}                     & Axitinib + FOLFOX vs BEV+FOLFOX (\Checkmark) \\
NCT00460603  & -- \cite{infante2013axitinib}                     & Axitinib + BEV+FOLFOX vs BEV+FOLFOX (\Checkmark) \\
NCT01418222  & GO27827 \cite{bendell2017phase}                   & Onartuzumab + BEV+FOLFOX vs BEV+FOLFOX (\Checkmark) \\
NCT01399684  & -- \cite{garcia2017randomized}                    & Parsatuzumab + BEV+FOLFOX vs BEV+FOLFOX (\Checkmark) \\
NCT00677443  & -- \cite{hong2012s}                  & SOX vs CAPOX (\Checkmark) \\
NCT00469443  & -- \cite{souglakos2012randomised}                 & BEV+FOLFIRI vs BEV+CAPIRI (\Checkmark) \\
NCT00153998  & CELIM \cite{folprecht2014survival}                & CET+FOLFIRI vs CET+FOLFOX (\Checkmark) \\
NCT01765582  & STEAM \cite{hurwitz2019phase}                     & BEV + FOLFOXIRI vs BEV + FOLFOX (\Checkmark) \\
--           & CAPIRI GOIM \cite{ciardiello2016cetuximab}        & CET+FOLFOX vs FOLFOX (\Checkmark) \\
NCT00636610  & Vismodegib trial \cite{berlin2013randomized}      & VIS+BEV+FOLFOX/FOLFIRI vs BEV+FOLFOX/FOLFIRI (\XSolidBrush) \\
UMIN000003253& FLEET \cite{soda2015multicenter}                  & CET + FOLFOX vs CET + XELOX (\XSolidBrush) \\
\bottomrule
\end{tabularx}
\begin{tablenotes}
\footnotesize
\item[*] The studies' hazard ratios are estimated by SynthIPD or IPDfromKM.
\item Drug acronyms: FOLFOX = leucovorin + 5-FU + oxaliplatin;
CAPOX = capecitabine + oxaliplatin; BEV = bevacizumab;
FOLFOXIRI = FOLFOX + irinotecan; IFL = irinotecan + bolus 5-FU + leucovorin;
FOLFIRI = leucovorin + 5-FU + irinotecan; XELIRI (CAPIRI) =
capecitabine + irinotecan; CET = cetuximab; SOX = S-1 + oxaliplatin;
VIS = Vismodegib.
\end{tablenotes}
\end{table}

\subsection{Clinical interpretation and risk-of-bias assessment}\label{app:nma-clinical-risk-bias}

\begin{figure}
    \centering
    \includegraphics[width=0.85\linewidth]{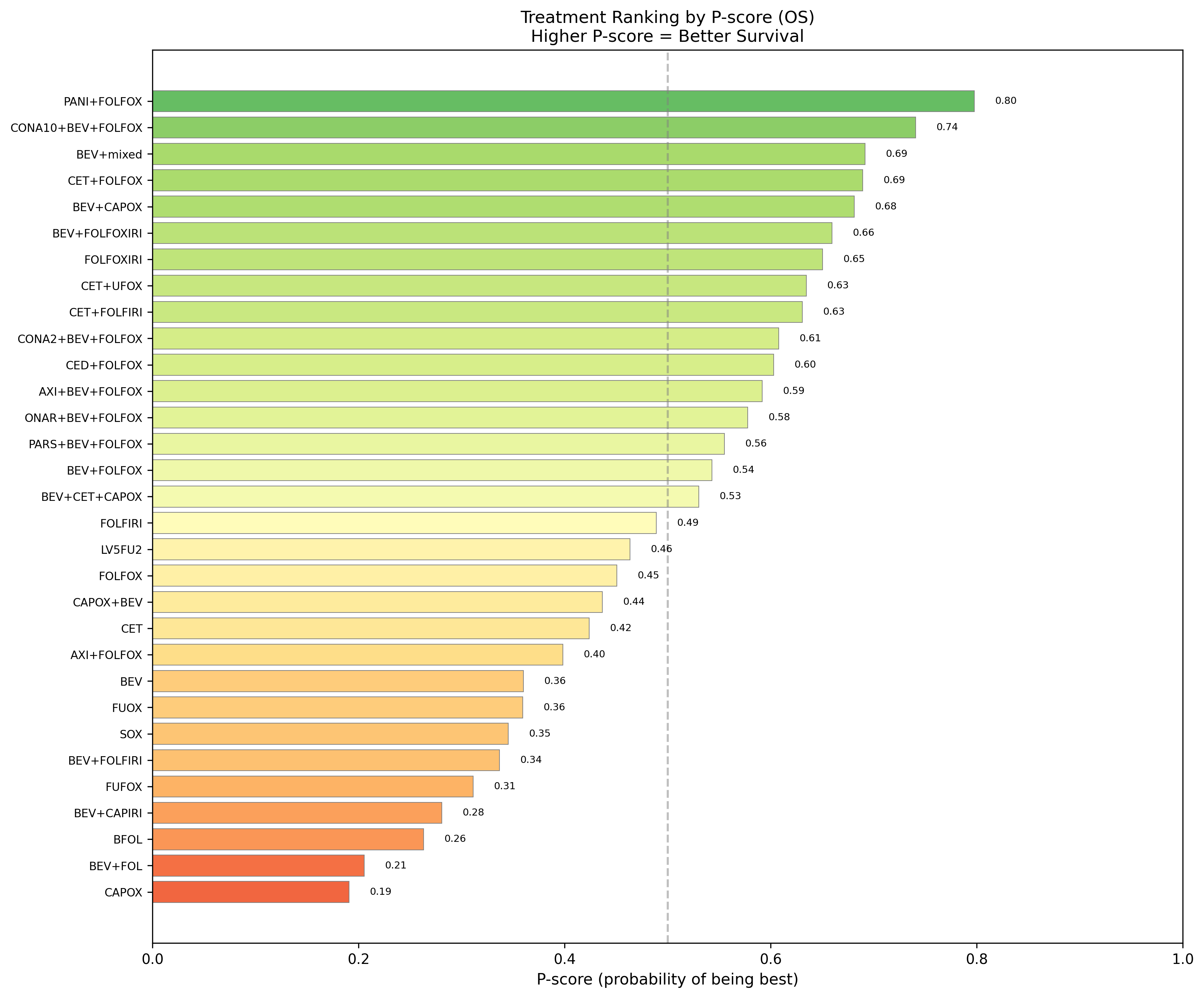}
    \caption{The probability of being best treatment arm (P-score) of all arms considered by the network meta analysis. BEV+mixed = BEV + CAPOX/FUOX/FOLFIRI/FUIRI in Figure \ref{fig:network_geometry}.}
    \label{fig:p score plot nma}
\end{figure}

Based on P-scores (Figure~\ref{fig:p score plot nma}), Panitumumab+FOLFOX achieved the highest ranking for OS. The rankings also highlight the benefit of incorporating bevacizumab, as most bevacizumab-based combinations appeared in the upper half of P-scores. In addition, using Bev+FOLFOXIRI as the reference, the estimated HRs (95\% CIs) were 1.09 (0.83, 1.44) for Bev+FOLFOX and 1.29 (0.94, 1.78) for Bev+FOLFIRI, respectively, suggesting outcomes favored Bev+FOLFOXIRI; these differences appeared more pronounced than those reported in~\cite{xu2021network}. Notably, several top-ranked regimens were evaluated in biomarker-defined populations; for example, trials of panitumumab- or cetuximab-based regimens often restricted enrollment to molecularly selected subgroups (e.g., RAS wild-type tumors). Because our screening criteria does not separate these subgroups from unselected populations, the corresponding P-scores may be inflated. Nevertheless, the overall pattern supports bevacizumab-based combinations, including CAPOX and FOLFOXIRI, as broadly effective options in this setting.


\begin{figure}
    \centering
    \includegraphics[width=0.95\linewidth]{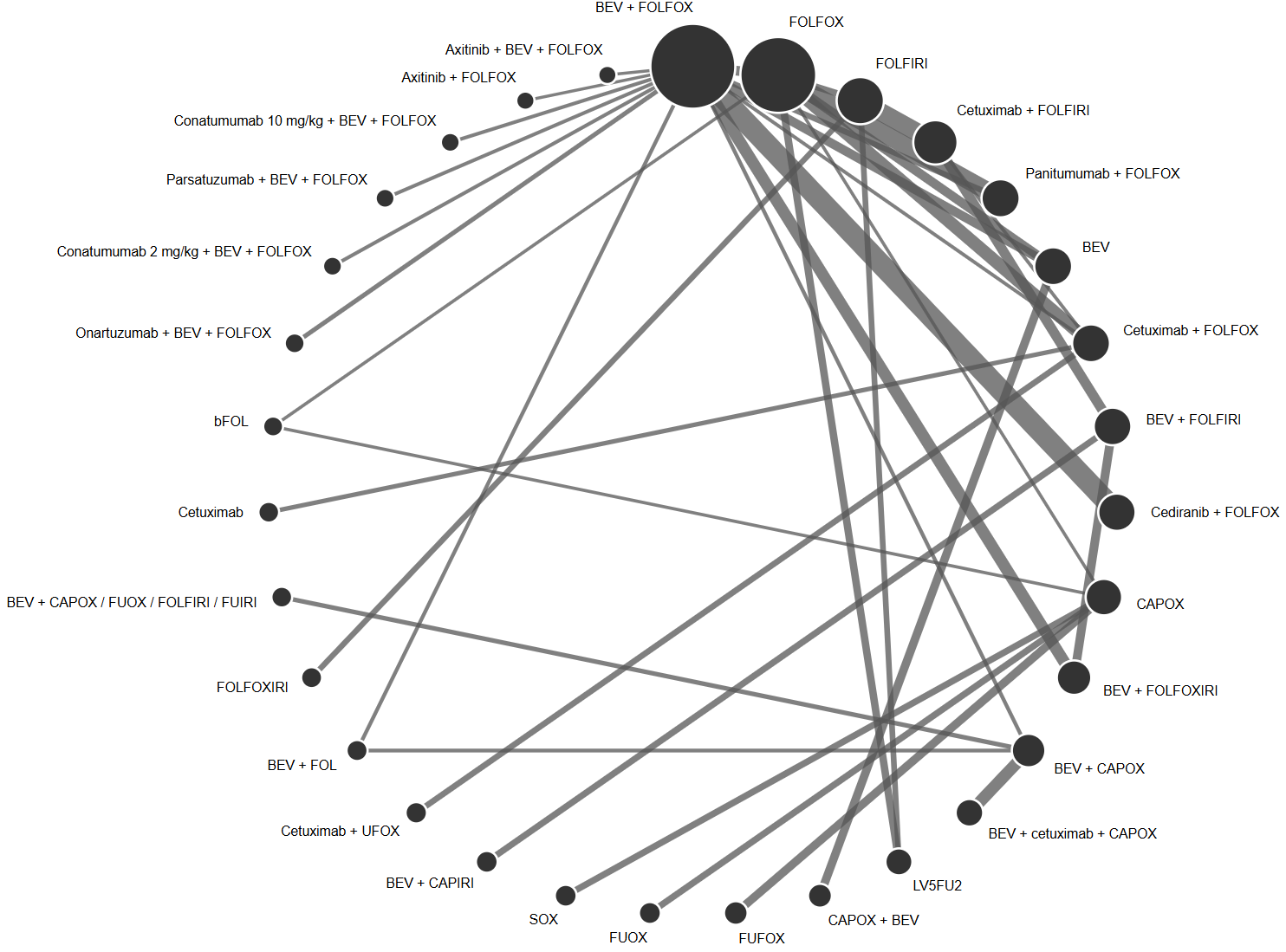}
    \caption{Network geometry plot for the included trials. Solid circles are proportional to total sample size. Line widths are proportional to each trial's (each comparison's) sample size. For the explanation of drug acronyms, see Tables \ref{tab:eligible trial meta analysis}, \ref{tab:eligible trial meta analysis2}. }
    \label{fig:network_geometry}
\end{figure}

\begin{figure}[!t]
    \centering
    \includegraphics[width=.5\linewidth]{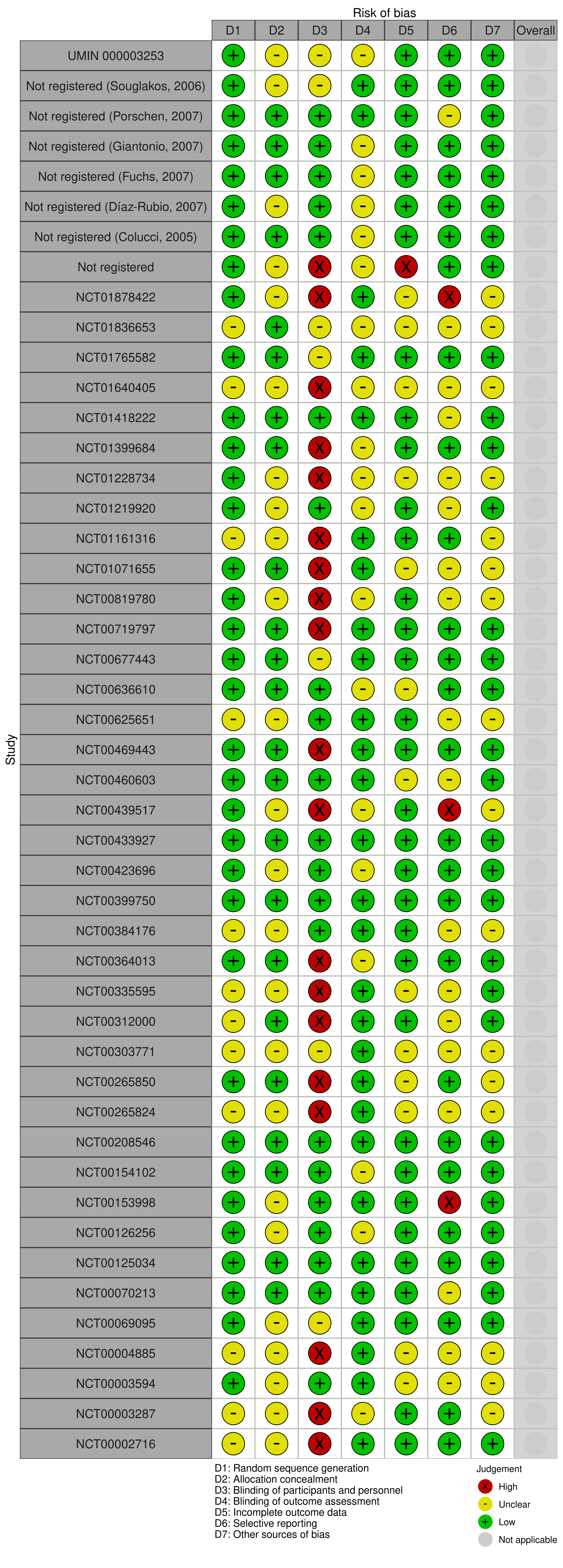}
    \caption{Risk of Bias table for network meta analysis.}
    \label{fig:rob nma}
\end{figure}

The risk-of-bias assessment results for all studies are presented in Figure~\ref{fig:rob nma}. Risk of bias was evaluated across seven domains: whether the study has risk on (1) randomization; (2) concealed allocation; (3) blinding of participants or not; (4) blinding of outcome assessment or not; (5) incomplete outcome data; (6) selective reporting (7) Other bias. Overall, the newly screened studies also show a low risk of bias. Note that all studies underwent risk of bias screening while a subset of them are included in the final Network geometry plot.

\clearpage
\bibliographystyle{plainnat}
\bibliography{ref}

\end{document}